\documentclass[notitlepage,aps,prl,reprint,twocolumn,longbibliography,superscriptaddress]
{revtex4-2}
\usepackage{graphicx}
\usepackage{amsmath}
\usepackage{amssymb}
\usepackage{comment}
\usepackage[colorlinks, allcolors=blue]{hyperref}
\usepackage[all]{hypcap}
\usepackage[mathlines]{lineno}
\usepackage{physics}
\usepackage{wrapfig}
\usepackage{lipsum}
\usepackage[normalem]{ulem}
\usepackage{siunitx}

\newcommand{\pref}[2]{\hyperref[#1]{\ref{#1}(#2)}}

\begin{document}
	\title{Quantum walks and correlated dynamics in an interacting synthetic Rydberg lattice}
	\author{Tao Chen}
\affiliation{Department of Physics, University of Illinois at Urbana-Champaign, Urbana, IL 61801-3080, USA}
    \author{Chenxi Huang}
\affiliation{Department of Physics, University of Illinois at Urbana-Champaign, Urbana, IL 61801-3080, USA}
	\author{Bryce Gadway}
\email{bgadway@psu.edu}
	\affiliation{Department of Physics, University of Illinois at Urbana-Champaign, Urbana, IL 61801-3080, USA}
 \affiliation{Department of Physics, The Pennsylvania State University, University Park, Pennsylvania 16802, USA}
 	\author{Jacob P. Covey}
	\email{jcovey@illinois.edu}
 \affiliation{Department of Physics, University of Illinois at Urbana-Champaign, Urbana, IL 61801-3080, USA}
	\date{\today}
\begin{abstract}
Coherent dynamics of interacting quantum particles plays a central role in the study of strongly correlated quantum matter and the pursuit of quantum information processors. Here, we present the state-space of interacting Rydberg atoms as a synthetic landscape on which to control and observe coherent and correlated dynamics. With full control of the coupling strengths and energy offsets between the pairs of sites in a nine-site synthetic lattice, we realize quantum walks, Bloch oscillations, and dynamics in an Escher-type ``continuous staircase". In the interacting regime, we observe correlated quantum walks, Bloch oscillations, and confinement of particle pairs. Additionally, we simultaneously tilt our lattice both up and down to achieve coherent pair oscillations. When combined with a few straightforward upgrades, this work establishes synthetic Rydberg lattices of interacting atom arrays as a promising platform for programmable quantum many-body dynamics with access to features that are difficult to realize in real-space lattices.
\end{abstract}

\maketitle

Quantum walks describe the propagation of quantum particles on periodic potentials, and are the quantum mechanical analog of classical random walks. They play an important role in quantum simulation~\cite{Mohseni2008,Plenio2008,Kitagawa-2010}, quantum search algorithms~\cite{Childs2003,Childs2004}, and even universal quantum computation~\cite{Childs2009,Venegas2012}. Moreover, quantum walks offer a way to benchmark the coherence and entanglement of interacting dynamics in programmable quantum devices~\cite{Mark2023}. Quantum walks have been studied with neutral atoms in optical lattices~\cite{Preiss-QWalk,Young2022}, superconducting circuits~\cite{Gong2021}, trapped ions~\cite{Schmitz2009,Zahringer2010}, photonics~\cite{Peruzzo2010}, and other platforms~\cite{Manouchehri2014}. These systems have demonstrated coherent evolution over tens of sites, where the geometry of the walks are constrained by the geometry of the underlying physical systems. The addition of a potential energy gradient leads to the observation of Bloch oscillations (BOs)~\cite{Dahan1996,Preiss-QWalk,Geiger2018} in which the particles remain localized and undergo periodic oscillations within nearby sites. Accordingly, a crossover between ballistic spreading and localized breathing dynamics is controlled by the relative strength of the kinetic versus potential energy. Coherent control over this crossover in scalable quantum systems is an important prerequisite for many quantum simulation and information paradigms.

Synthetic dimensions provide an alternative bottom-up approach for Hamiltonian engineering to explore quantum walks and coherent dynamics of quantum particles~\cite{Mancini2015,an2018corr,Yuan:18,SundarPRA,Ozawa2019,Kanungo2022,Oliver2023,Englebert2023,Deng2022,Parto2023}. Instead of tunneling between real-space sites such as the anti-nodes of an optical lattice or the transmons in a superconducting circuit, the ``particle"  hops within the state space of a host quantum system. The hopping rates between each pair of sites in the synthetic dimension is controlled with a drive that couples the two states. The local energy offset depends on the detuning of that drive. This paradigm offers several new opportunities: First, the ``connectivity" between each pair of sites is fully programmable, enabling an emergent geometry and even the addition of topological phases~\cite{Ozawa2019,chen2024}. Second, the energy landscape is also fully programmable, which allows for the addition of any disorder pattern, lattice tilts~\cite{Oliver2023,Englebert2023}, or non-Euclidean geometry such as in the Escher-type ``impossible" continuous stairs~\cite{Mueller-Escher}. 

Here we utilize the large state space of individual Rydberg atoms to encode a synthetic lattice~\cite{Kanungo2022,chen2024,Lu2024,chen2024caging} with nine sites. Tunneling amplitudes and energy offsets are controlled with ``nearest neighbor" drives, while interactions between the particles arise from the Rydberg dipolar exchange. Starting with single particles, we demonstrate highly coherent quantum walks in a flat lattice, Bloch oscillations in a tilted lattice, and dynamics within an Escher-type ``continuous staircase" with a periodic boundary condition (PBC) that admits interference between clockwise and counterclockwise pathways.
We then introduce interactions by preparing a pair of closely-spaced Rydberg atoms, where we study the interplay of interactions and lattice tilt on the dynamics. We observe that interaction strengths comparable to the lattice tilt induce free transport by a breakdown of Stark localization~\cite{Mendez-stark}, but that interactions larger than the lattice tilt lead to re-entrant localization. Finally, we explore the scenario where the lattice is tilted both up \textit{and} down simultaneously, observing the Floquet control of correlated pair hopping. Straightforward upgrades to our system could increase the state space and tunneling rates by an order of magnitude, enable two-dimensional geometries, and extend to hundreds of atoms. %\textcolor{blue}{while mitigating positional disorder}. 
This work demonstrates the coherence and programmability of the Rydberg atom state space as a new platform for quantum simulation, quantum search algorithms, and even universal quantum computing.

\begin{figure*}[t]
	\includegraphics[width=\textwidth]{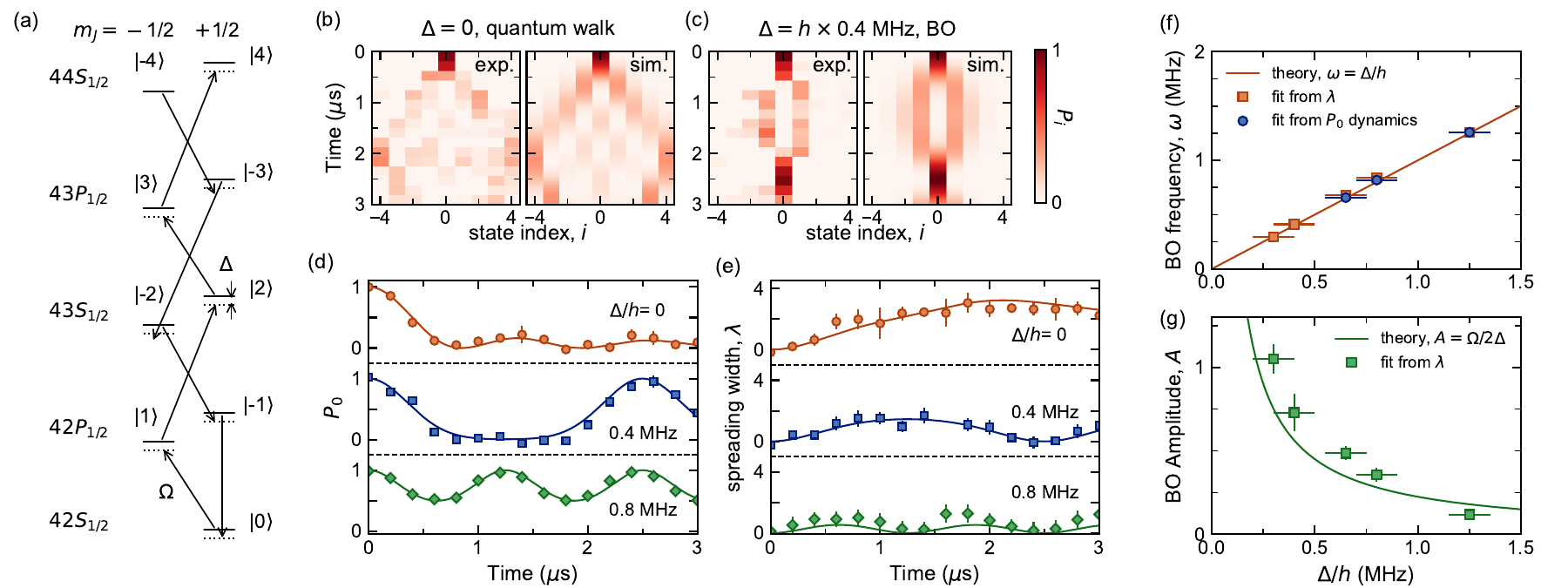}
	\caption{\textbf{Bloch oscillations in a synthetic Rydberg state lattice.}
        \textbf{(a)}~Level structure used to implement a 9-site Rydberg state lattice. Each state pair is coupled by a corresponding microwave tone (black arrows) with a Rabi rate $\Omega$ and a detuning of $\Delta$. 
        \textbf{(b,c)}~Quantum walk [(b), $\Delta=0$] and BO [(c), $\Delta/h = 0.4(1)$~MHz] dynamics in a 9-site lattice with $\Omega/h = 0.45(1)$~MHz. 
        \textbf{(d,e)}~Time evolution of the population in the initial $|0\rangle$ state, $P_0$, and the wavepacket spreading width $\lambda$ for different detunings (see text).
        \textbf{(f,g)}~BO frequency $\omega$ and amplitude $A$ as functions of detuning $\Delta$. Solid lines indicate the theory predictions. The error bars in (d,e) relate to the standard error of multiple independent data sets. The vertical (horizontal) error bars in (f,g) relate to the standard error of the fits (uncertainty of the 
        calibrated detunings).
}
\label{FIG:fig1}
\end{figure*}  

\textit{Single particle dynamics}.--- Our synthetic lattice is routinely implemented in a probabilistically loaded one-dimensional tweezer array of single $^{39}$K atoms with a series of dimerized trap configurations \cite{JacksonPRR,chen2024,chen2024caging}. We postselect on having only a single trap loaded in each trap dimer to observe the non-interacting single particle dynamics. The trapped atoms are globally excited to Rydberg state $\ket{0}=\ket{42S_{1/2},m_J=1/2}$. We then couple it to eight others -- four on each side -- with microwave drives~\cite{Kanungo2022,chen2024} such that the ``particle" is initialized in the middle of the lattice chain [see Fig.~\ref{FIG:fig1}(a)]. In this work, we primarily detect the initial Rydberg state, $|0\rangle$, or the state on one of the ends, $|4\rangle$, by mapping it back to the ground state such that it is bright to subsequent fluorescence detection of the atom. However, we are also able to detect each site independently, which we use to track the full single-particle dynamics in the synthetic lattice. We note that our experimental data throughout has been re-scaled based on the known preparation and measurement errors~\cite{chen2024,chen2024caging,SuppMats}.

We first investigate
wavepacket spreading under different tilts,
described by the one-atom Hamiltonian
\begin{equation}\label{single-particle}
 H_{\rm sp} = \Delta\sum_{j} j c^\dagger_j c_j + \frac{\Omega}{2}\sum_{j} \left(c^\dagger_{j+1}c_{j} + {\rm h.c.}\right)
\end{equation}
with the hopping rates $\Omega/2$ between each adjacent state pair (kinetic energy) and the programmable on-site potential energy $j \times \Delta$ (most generally $\Delta_j$). Figures~\pref{FIG:fig1}{b,c} track the time evolution of the populations in all nine sites, showing both experimental results and simulations, respectively for the flat lattice case ($\Delta=0$) and for a tilted lattice ($\Delta/h=0.4$ MHz; $\Delta\approx\Omega$). Our observations highly agree with the numerical simulations based on Eq.(\ref{single-particle}) with no free parameters. For the flat lattice case the wavepacket follows continuous-time quantum walk with reflections from the open boundaries, while in a tilted lattice we see the breathing-mode feature of the BO \cite{Preiss-QWalk}, i.e., the wavepacket primarily oscillates between the initial center site and nearby neighbors. 

As shown in Figs.~\pref{FIG:fig1}{d,e}, the manifestations of such ballistic spreading and tilt-dependent localized breathing dynamics are effectively explicable via the time evolutions of both the center-site population, $P_0$, and the width of the wavepacket across the full lattice, $\lambda = \sum_{j} |j| P_j$. We fit the measured wavepacket spreading width under lattice tilts with $\lambda(t) = A [1- \cos{(2\pi\omega t)}]$ \cite{Price-Shaking}. The fitting results of BO amplitude and frequency, illustrated in Figs.~\pref{FIG:fig1}{f,g}, align well with the theoretical predictions $\omega=\Delta/h$ and $A=\Omega/(2\Delta)$.
For $\Delta >\Omega$, the oscillation frequency can also be resolved from the measured $P_0$ dynamics, which we employ subsequently to describe the oscillating behaviors of an interacting pair.

\begin{figure}[t]
	\includegraphics[width=0.5\textwidth]{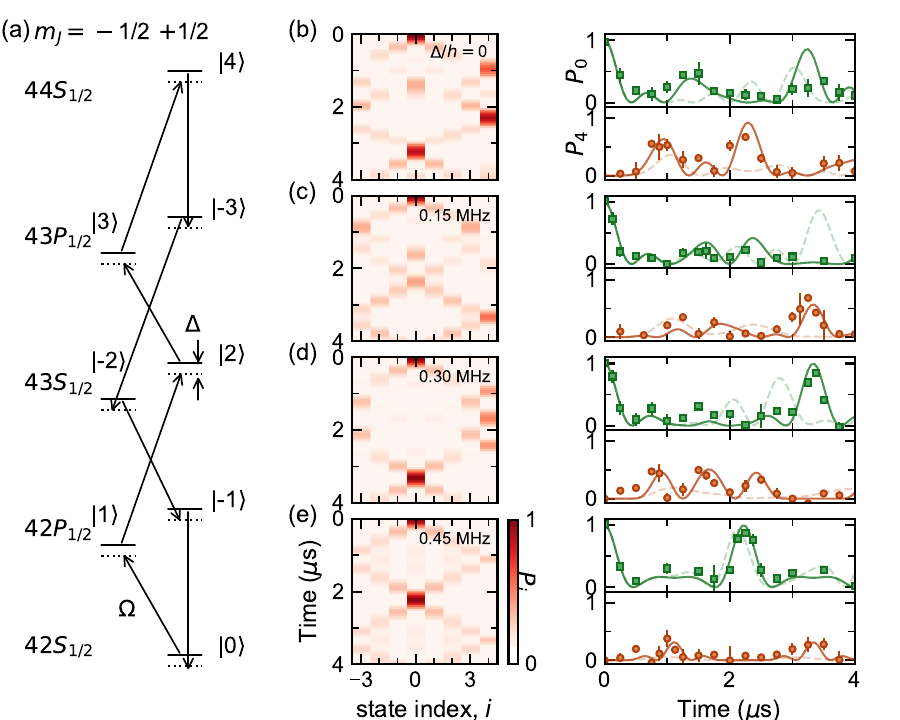}
	\caption{\textbf{Dynamics for a non-interacting 8-site Rydberg lattice with Escher structure.}
        \textbf{(a)}~State configuration to implement the continuous staircase. Here we use $\Omega/h=0.90(2)~{\rm MHz}$.
        \textbf{(b-e)}~Left panels: simulated population dynamics for different detunings $\Delta/h = \{0, 0.15, 0.30, 0.45\}~{\rm MHz}$. Right panels: measured $P_0$ (green squares) and $P_4$ (orange circles) dynamics for corresponding detunings in left panels. Solid lines are the same simulations as those in the left panels, while the dashed lines, for comparison, indicate the population dynamics under open boundary condition by disconnecting the $\ket{4} \to \ket{-3}$ transition. The error bars relate to the standard errors of multiple independent data sets.        
}
\label{FIG:fig2}
\end{figure}

Next, we build a ring geometry comprising eight states under PBCs, with the single atom initialized in $\ket{0}$; see Fig.~\pref{FIG:fig2}{a}. Finite detunings lead to an Escher-type ``continuous staircase" configuration~\cite{Mueller-Escher}, wherein the energy cost is always positive (negative) when moving clockwise (counterclockwise). Similar to electromagnetic induction, here the potential gradient (synthetic electric field) can be viewed as arising from a time-varying flux that pierces the ring lattice. The inescapability of residual time dependence in the effective Hamiltonian is a hallmark of the Escher staircase scenario~\cite{Mueller-Escher}. Still, as for open boundaries, as the step height of the continuous staircase increases, the wavepacket dynamics again witness a crossover from ballistic spreading [$\Delta=0$, Fig.~\pref{FIG:fig2}{b}] to Bloch oscillation [$\Delta=\Omega/2$, Fig.~\pref{FIG:fig2}{e}]. 

The primary distinguishing feature between these two regimes is the transient refocusing observed at both the initial site $\ket{0}$ and the opposing site $\ket{4}$, due to the interference of the two pathways in clockwise and counterclockwise directions. For the flat lattice case in Fig.~\pref{FIG:fig2}{b}, the wavepackets spreading in two directions accumulate identical finite phases, resulting in significant refocusing in both $\ket{0}$ and $\ket{4}$. However, when introducing a finite small detuning, e.g., $\Delta=\Omega/6$ in Fig.~\pref{FIG:fig2}{c}, we attribute the weak refocusing (projections on $\ket{0}$ and $\ket{4}$) in the $4~\mu{\rm s}$ window to the destructive interference, as the two pathways generate different phases due to the opposite-sign energy costs. With larger tilts, $\Delta=\Omega/3$ and $\Omega/2$ in Figs.~\pref{FIG:fig2}{d,e} respectively, almost perfect refocusing in the initial $\ket{0}$ is observed. The refocusing time $t_r$ can be determined by simply letting the accumulated phase $\Delta t_r/\hbar = 2\pi$, leading to $\sim 3.3~\mu{\rm s}$ and $\sim 2.2~\mu{\rm s}$ for the two cases respectively as evidenced by our experimental observation. Here $t_r$ can be regarded as the BO period, in agreement with the theoretically predicted oscillating frequency $\omega=\Delta/h$ \cite{Price-Shaking}. 
In this exotic scenario of a periodic boundary lattice with a continuous tilt, the Bloch period can be sufficiently long that the wavepacket fully wraps around the lattice before beginning to refocus~\cite{Mueller-Escher}.

By comparing to the dynamics for open boundaries (dashed curves), we note from Fig.~\pref{FIG:fig2}{e} that the boundary conditions don't impact the dynamics once the tilt is sufficiently large such that the BO amplitude is smaller than the system size.Considering that the dynamics reflects the interference of right- and left-moving paths, we note that the global flux piercing the ring for PBCs can also play an important role~\cite{chen2024,chen2024caging}. The results of Fig.~\ref{FIG:fig2} are for an initial flux of zero, calibrated based on the population dynamics for $\Delta=0$ and our control of tunneling phases~\cite{SuppMats}. Control of this flux, along with programmable potentials and interactions, could enable future investigations of spectral topology under PBCs~\cite{Thouless-Twist}.

\textit{Interacting atom pairs}.--- We now 
discuss interacting dynamics and pair hopping
by utilizing a pair of closely-spaced Rydberg atoms labelled as $A$ and $B$ with a spatial separation $d_{AB}$. The interacting Hamiltonian is 
\begin{equation}
 H_{\rm int} = \sum_{i,j} V_{ij} c^\dagger_{i,A} c^\dagger_{j, B} c^{}_{j,A} c^{}_{i, B} + {\rm h.c.},
\end{equation}
with $V_{ij} \propto C_3^{ij}/d_{AB}^3$ for the dipolar exchange $\ket{i}_A\ket{j}_B \leftrightarrow \ket{j}_A\ket{i}_B$. We scale all $V_{ij}$ to $V=V_{0,-1}$ with the calculated $C_3$ coefficients. In the experiment we vary the interaction strength by changing the tweezer separation $d_{AB}$. Here we return to the 9-state synthetic lattice in Fig.~\ref{FIG:fig1}, and each atom also experiences the single-particle Hamiltonian (\ref{single-particle}). In our numerical simulations, the largely detuned state-changing interaction terms are excluded as we work in relatively weak interaction regime \cite{SuppMats}.

\begin{figure}[]
	\includegraphics[width=\columnwidth]{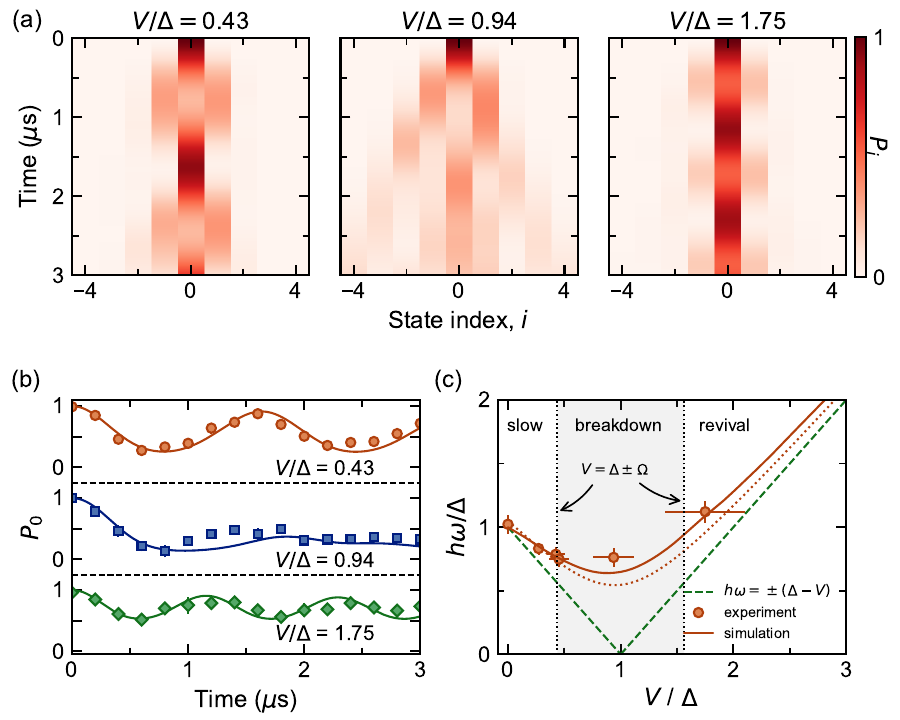}
	\caption{\textbf{BO in an interacting atom pair.}
        \textbf{(a)}~Simulated averaged population dynamics for atom pair in an interacting Rydberg state lattice under different interaction strength $V/\Delta = \{0.43, 0.94, 1.75\}$ with $\Delta/h=0.8~{\rm MHz}$ and $\Omega/h=0.45~{\rm MHz}$.
        \textbf{(b)}~Time evolution of $P_0$ under different interaction strength. Solid lines give the ideal numerical simulation results.
        The errorbars show the standard errors from multiple independent datasets.
        \textbf{(c)}~Oscillation frequency $\omega$ v.s. interation-to-detuning ratio $V/\Delta$. With an increase of $V/\Delta$, the oscillation first slows down, then gets breakdown but finally revival. The boundaries of the breakdown region are determined by $V=\Delta \pm \Omega$, wherein damping feature shows up for short-time $P_0$ dynamics. The solid line is obtained by fitting the numerically simulated ideal $P_0$ dynamics. The dashed line with $h\omega = \pm(\Delta - V)$ gives the energy gap between pair state $\ket{0,0}$ and triplets $(\ket{\pm 1,0} + \ket{0,\pm 1})/\sqrt{2}$ in $\Omega \to 0$ limit. The dotted line is the energy gap modified by finite small $\Omega$ in formula $h\omega \approx \sqrt{|\Delta - V|^2+\Omega^2}$ for 3-state system $\{\ket{0},\ket{\pm1}\}$ under interaction \cite{SuppMats}. In experiment, we use $\Delta/h = 0.8(1)~{\rm MHz}$, $\Omega/h = 0.45(1)~{\rm MHz}$. The vertical (horizontal) error bars come from fittings to the experimental datasets (uncertainties of both $V$ and $\Delta$).
}
\label{FIG:fig3}
\end{figure}

We consider the case of a highly tilted lattice with $\Delta/h=0.8(1)$ MHz.
Our analysis reveals significant deviations from the non-interacting case. 
Figure~\pref{FIG:fig3}{a} shows the simulated time evolution of the averaged population in each site, $P_i=(\langle c_{i,A}^\dagger c_{i,A}\otimes I_B\rangle + \langle I_A\otimes c_{i,B}^\dagger c_{i,B}\rangle)/2$, under different interaction strength. For small $V<\Delta$ that perturbatively changes the effective tilt, we still anticipate observing BOs but with modified $V$-dependent frequencies. Conversely, strong interactions ($V>\Delta$) lead to confinement, relating to frozen bound pairs~\cite{chen2024} with small hopping probability between $\ket{0,0}$ and adjacent triplet states.
In the intermediate regime where $V\approx\Delta$, we expect to observe delocalization in an anti-correlated quantum walk of the $A$ and $B$ atoms~\cite{SuppMats}. 
Experimentally, we analyze the dynamics of $P_0$ to characterize the oscillating behavior; see Fig.~\pref{FIG:fig3}{b}. We observe small-amplitude oscillations with a frequency different from that for the non-interacting case, for both weak and strong $V$. However, a notable damping phenomenon, devoid of perfect revival, manifests when the interaction roughly compensates the detuning of pair state coupling, indicating the breakdown of BOs.

Based on our observations, numerics, and energy arguments~\cite{SuppMats}, we 
define the
``localization breakdown'' region with interaction-induced transport as
$\Delta - \Omega < V < \Delta + \Omega$. 
From fittings of $P_0$ dynamics with a damped sine
function, the damping coefficients are almost zero outside of this breakdown region \cite{SuppMats}, while the oscillating frequency first decreases then ramps up as $V$ increases [see Fig.~\pref{FIG:fig3}{c}]. 
We can quantitatively relate this frequency to the energy gap between the $|0,0\rangle$ and the $(|\pm1,0\rangle+|0,\pm1\rangle)/\sqrt{2}$ triplet states. For $\Omega\rightarrow0$, this is given by $\pm(\Delta-V)$~\cite{SuppMats}. Considering small but finite $\Omega$, the energy gaps of the truncated lattice ($\ket{-1}$, $\ket{0}$, and $\ket{1}$) are refined to $\sim$$\sqrt{|\Delta-V|^2+\Omega^2}$. This latter form agrees well with our observations outside of the breakdown region. We note a slight deviation from the frequency based on numerics in the large $V$ limit, which we attribute to the role of outer sites ($|j| > 1$). Overall, these observations illustrate how correlated quantum walks and Bloch oscillations can be used to probe the effective band structure of the interacting system.

\begin{figure}[t!]
	\includegraphics[width=\columnwidth]{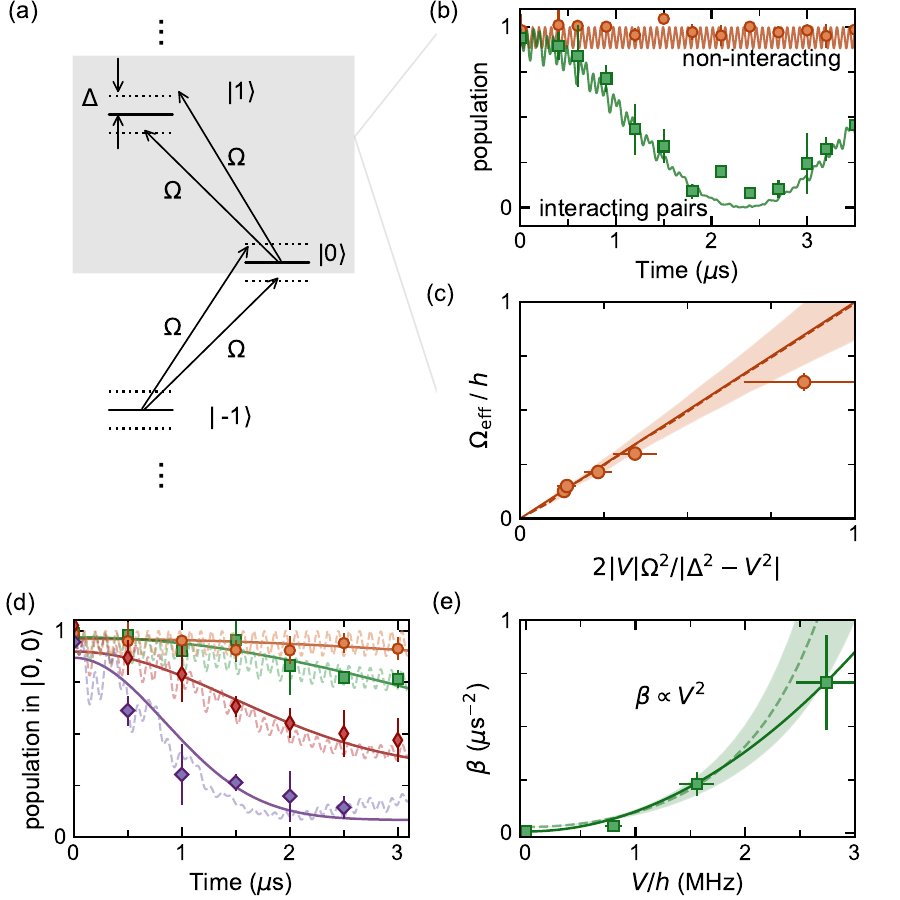}
	\caption{\textbf{Pair hopping and pair dynamics under bichromatic driving.}
        \textbf{(a)}~The driving scheme. Each Rydberg state pair is coupled by two equal strength ($\Omega$) microwave tones, respectively detuned from resonance by $\pm \Delta$. In (b,c), we study only the two-state system of $\ket{0}$ and $\ket{1}$  (gray box) under a bichromatic drive.
        \textbf{(b)}~Time evolution of the populations in $\ket{0}$ for single atoms (orange circles) and $\ket{0,0}$ for interacting pairs with $V/h= -1.56(9)~{\rm MHz}$ (green squares). Here $\Omega/h = 1.92(4)~{\rm MHz}$ and $\Delta/h = 7.2(1)~{\rm MHz}$. The solid lines are numerical simulations. 
        \textbf{(c)}~Comparison of the experimentally measured pair-hopping ($\ket{0,0}\leftrightarrow\ket{1,1}$) rate $\Omega_{\rm eff}$ to the perturbation theory prediction (solid line). The dashed line shows the result from fittings to the numerically calculated $P_{\ket{00}}$ dynamics.
        \textbf{(d)}~Time evolution of the SPAM-corrected $\ket{0,0}$ population for the 9-site lattice under bichromatic driving of each nearest-neighbor transition. The data plots, from top to bottom, relate to interactions $V/h = \{0, 0.80(5), 1.56(9), 2.70(16)\}~{\rm MHz}$ with $\Omega/h=0.90(2)~{\rm MHz}$ and $\Delta/h=5.0(1)~{\rm MHz}$. Solid lines are exponential fits $P_{\ket{0,0}}(t) = a + b\exp(-\beta t^2)$ to the experimental data, while
        dashed lines are numerical simulations.
        \textbf{(e)}~The damping coefficient $\beta$ vs. the interaction strength. The dashed line shows the fit of short-time numerics [dashed lines in (d)] to the same form. The solid line is a polynomial fit to the experimental data, showing the scale $\beta\propto V^2$. The shaded regions in (c,e) result from accounting for
        known parameter uncertainties.
        The vertical error bars in (b,d) are standard errors from multiple independent measurements, while in (c,e) they are
        standard errors of the fit. The horizontal errorbars in (c,e) reflect
        uncertainties of the calibrated values of $V$, $\Omega$, and $\Delta$.
}
\label{FIG:fig4}
\end{figure}

Finally, we explore a unique capability of synthetic lattices by simultaneously tilting the lattice both upward and downward. This scenario corresponds to a bichromatic drive with detunings $\pm\Delta$ for each pair of states; see Fig.~\ref{FIG:fig4}. We explore the interplay of this bichromatic drive and interactions $V$ between the atoms when $\Delta\approx4\Omega$. We start with a two-level system, $|0\rangle$ and $|1\rangle$, as illustrated by the gray box in Fig.~\pref{FIG:fig4}{a}. In the non-interacting case, this corresponds simply to highly detuned Rabi dynamics, with a nearly static $P_0 \approx 1$. However, when interactions with strength $V=0.2\Delta$ are added, we observe large-amplitude pair oscillations between $|0,0\rangle$ and $|1,1\rangle$. Measuring for different $V$ values~[see Fig.~\pref{FIG:fig4}{c}], we find good agreement with the expected rate of activated pair hopping, $\Omega_\textrm{eff}=2|V|\Omega^2/|\Delta^2-V^2|$~\cite{SuppMats}.
This activated pair-hopping~\cite{Meinert-Floquet} is controlled independently from the hopping of singles, thus distinct from the hopping of bound pairs in flat lattices ($\Delta = 0$) when $V \gg \Omega$~\cite{Winkler2006,Preiss-QWalk,chen2024}.

This observation highlights an exciting opportunity that arises naturally in the synthetic dimension -- because the physical states have different energies and all hopping is activated by driving (similar to a strongly tilted lattice~\cite{anyon-hubb-theory,kwan2023realization}), 
multiple driving tones can be combined with strong interactions to introduce exotic higher-order processes.
In the simple two-state case, the observed pair-hopping can be viewed as the activation of ``flip-flip'' ($\sigma^+\sigma^+$) and ``flop-flop'' ($\sigma^-\sigma^-$) spin interactions~\cite{SuppMats}, relevant for the realization of Kitaev spin chains~\cite{Kitaev-DasSarma}. This process is also analogous to the recent Floquet activation~\cite{Huanqian-Floquet} of Rydberg anti-blockade~\cite{Antiblockade}.

With this understanding of the bichromatic drives for the two-level system, we now return to the full nine-level system shown in Fig.~\pref{FIG:fig1}{a}, but with bichromatic drives. We measure the evolution of the $|0,0\rangle$ population under different interaction strength $V$; see Fig.~\pref{FIG:fig4}{d}. In the limit of zero interaction, $P_{\ket{0,0}}$ remains high just as it did for the two-level case. As the interaction strength is increased, we find that $P_{00}$ decays with an increasing rate. The measured dynamics are in good agreement with simulations, which suggest an interaction-facilitated delocalization by pair-hopping
that appears as a Gaussian decay on the timescales relevant to
experiment~\cite{SuppMats}. By fitting to the expected Gaussian decay $e^{-\beta t^2}$, we find the damping coefficient $\beta$ scales as $\sim V^2$, in agreement with form expected from theory in the weak-interaction limit~\cite{SuppMats}; see Fig.~\pref{FIG:fig4}{e}.

\textit{Concluding discussion}.--- This work establishes synthetic Rydberg lattices in interacting atom arrays as a promising platform to explore programmable quantum many-body dynamics. By studying quantum walks for single particles and correlated pairs, we demonstrate coherent evolution of a system comparable in scale with that of other atomic platforms. %Additionally, 
We leverage the unique capabilities of synthetic dimensions to explore scenarios that are challenging to achieve in
real dimensions, such as arbitrary energy landscapes, simultaneous up-and-down tilting,
and non-Euclidean geometry such as the Escher-type ``continuous staircase''.
Realistic extensions will enable the study of exotic geometries such as tesseracts, M\"{o}bius strips, and Klein bottles.
Moreover, the varied and long-range interaction energies across our synthetic lattice provide another unique aspect of our system that may aid in the study of disordered many-body dynamics with complex interaction graphs~\cite{Zhang2023Painter}. Finally, although quantum walks with non-interacting bosons present a scenario for boson sampling~\cite{Young2024sample} and tests of quantum advantage~\cite{Zhong2020sample}, quantum walks and related dynamics with strongly- and long-range-interacting particles presents a much richer landscape. Indeed, programmable interactions set our work apart from photonic approaches.

We note several straightforward ways to advance the coherence and increase the timescales of dynamics in this platform. First, it would be beneficial to implement trapping of our Rydberg states, either by blue-detuned bottlebeam traps~\cite{Barredo2020} or with alkaline earth(-like) atoms with trappable ionic cores~\cite{Wilson2022}. Positional disorder in free flight due to finite temperature is the main limitation to coherent many-body dynamics. Even in the current system, however, this issue could be mitigated with Raman sideband cooling~\cite{Thompson13,kaufman12}, which would be beneficial also for trappable Rydberg atoms. Second, a larger magnetic field would offer a larger spacing between the desired states and other Zeeman states, which would reduce unwanted off-resonant couplings and enable faster dynamics. We anticipate than an order of magnitude improvement would be straightforward. Third, the use of higher Rydberg states or circular Rydberg states~\cite{Wu2023} -- potentially even in a cryogenic environment~\cite{Schymik2021} -- would mitigate the effect of the finite Rydberg state lifetime. Already, this work sets the stage for complex many-body Rydberg dynamics with a large degree of programmability that can enable the study of few-body and many-body localized~\cite{Abanin2019,Schreiber2015,Lukin2019,Yao2016,Morong2021} and topological~\cite{Rachel2018,Walter2023} phases.

We thank Tabor Electronics greatly for the use of an arbitrary waveform generator demo unit. This material is based upon work supported by the National Science Foundation under grant No.~1945031 and the AFOSR MURI program under agreement number FA9550-22-1-0339.

\bibliographystyle{apsrev4-2}
\bibliography{intBO}

\begin{thebibliography}{59}%
\makeatletter
\providecommand \@ifxundefined [1]{%
 \@ifx{#1\undefined}
}%
\providecommand \@ifnum [1]{%
 \ifnum #1\expandafter \@firstoftwo
 \else \expandafter \@secondoftwo
 \fi
}%
\providecommand \@ifx [1]{%
 \ifx #1\expandafter \@firstoftwo
 \else \expandafter \@secondoftwo
 \fi
}%
\providecommand \natexlab [1]{#1}%
\providecommand \enquote  [1]{``#1''}%
\providecommand \bibnamefont  [1]{#1}%
\providecommand \bibfnamefont [1]{#1}%
\providecommand \citenamefont [1]{#1}%
\providecommand \href@noop [0]{\@secondoftwo}%
\providecommand \href [0]{\begingroup \@sanitize@url \@href}%
\providecommand \@href[1]{\@@startlink{#1}\@@href}%
\providecommand \@@href[1]{\endgroup#1\@@endlink}%
\providecommand \@sanitize@url [0]{\catcode `\\12\catcode `\$12\catcode
  `\&12\catcode `\#12\catcode `\^12\catcode `\_12\catcode `\%12\relax}%
\providecommand \@@startlink[1]{}%
\providecommand \@@endlink[0]{}%
\providecommand \url  [0]{\begingroup\@sanitize@url \@url }%
\providecommand \@url [1]{\endgroup\@href {#1}{\urlprefix }}%
\providecommand \urlprefix  [0]{URL }%
\providecommand \Eprint [0]{\href }%
\providecommand \doibase [0]{https://doi.org/}%
\providecommand \selectlanguage [0]{\@gobble}%
\providecommand \bibinfo  [0]{\@secondoftwo}%
\providecommand \bibfield  [0]{\@secondoftwo}%
\providecommand \translation [1]{[#1]}%
\providecommand \BibitemOpen [0]{}%
\providecommand \bibitemStop [0]{}%
\providecommand \bibitemNoStop [0]{.\EOS\space}%
\providecommand \EOS [0]{\spacefactor3000\relax}%
\providecommand \BibitemShut  [1]{\csname bibitem#1\endcsname}%
\let\auto@bib@innerbib\@empty
%</preamble>
\bibitem [{\citenamefont {Mohseni}\ \emph {et~al.}(2008)\citenamefont
  {Mohseni}, \citenamefont {Rebentrost}, \citenamefont {Lloyd},\ and\
  \citenamefont {Aspuru-Guzik}}]{Mohseni2008}%
  \BibitemOpen
  \bibfield  {author} {\bibinfo {author} {\bibfnamefont {M.}~\bibnamefont
  {Mohseni}}, \bibinfo {author} {\bibfnamefont {P.}~\bibnamefont {Rebentrost}},
  \bibinfo {author} {\bibfnamefont {S.}~\bibnamefont {Lloyd}},\ and\ \bibinfo
  {author} {\bibfnamefont {A.}~\bibnamefont {Aspuru-Guzik}},\ }\href
  {https://doi.org/10.1063/1.3002335} {\bibfield  {journal} {\bibinfo
  {journal} {The Journal of Chemical Physics}\ }\textbf {\bibinfo {volume}
  {129}},\ \bibinfo {pages} {174106} (\bibinfo {year} {2008})}\BibitemShut
  {NoStop}%
\bibitem [{\citenamefont {Plenio}\ and\ \citenamefont
  {Huelga}(2008)}]{Plenio2008}%
  \BibitemOpen
  \bibfield  {author} {\bibinfo {author} {\bibfnamefont {M.~B.}\ \bibnamefont
  {Plenio}}\ and\ \bibinfo {author} {\bibfnamefont {S.~F.}\ \bibnamefont
  {Huelga}},\ }\href {https://doi.org/10.1088/1367-2630/10/11/113019}
  {\bibfield  {journal} {\bibinfo  {journal} {New Journal of Physics}\ }\textbf
  {\bibinfo {volume} {10}},\ \bibinfo {pages} {113019} (\bibinfo {year}
  {2008})}\BibitemShut {NoStop}%
\bibitem [{\citenamefont {Kitagawa}\ \emph {et~al.}(2010)\citenamefont
  {Kitagawa}, \citenamefont {Rudner}, \citenamefont {Berg},\ and\ \citenamefont
  {Demler}}]{Kitagawa-2010}%
  \BibitemOpen
  \bibfield  {author} {\bibinfo {author} {\bibfnamefont {T.}~\bibnamefont
  {Kitagawa}}, \bibinfo {author} {\bibfnamefont {M.~S.}\ \bibnamefont
  {Rudner}}, \bibinfo {author} {\bibfnamefont {E.}~\bibnamefont {Berg}},\ and\
  \bibinfo {author} {\bibfnamefont {E.}~\bibnamefont {Demler}},\ }\href
  {https://doi.org/10.1103/PhysRevA.82.033429} {\bibfield  {journal} {\bibinfo
  {journal} {Phys. Rev. A}\ }\textbf {\bibinfo {volume} {82}},\ \bibinfo
  {pages} {033429} (\bibinfo {year} {2010})}\BibitemShut {NoStop}%
\bibitem [{\citenamefont {Childs}\ \emph {et~al.}(2003)\citenamefont {Childs},
  \citenamefont {Cleve}, \citenamefont {Deotto}, \citenamefont {Farhi},
  \citenamefont {Gutmann},\ and\ \citenamefont {Spielman}}]{Childs2003}%
  \BibitemOpen
  \bibfield  {author} {\bibinfo {author} {\bibfnamefont {A.~M.}\ \bibnamefont
  {Childs}}, \bibinfo {author} {\bibfnamefont {R.}~\bibnamefont {Cleve}},
  \bibinfo {author} {\bibfnamefont {E.}~\bibnamefont {Deotto}}, \bibinfo
  {author} {\bibfnamefont {E.}~\bibnamefont {Farhi}}, \bibinfo {author}
  {\bibfnamefont {S.}~\bibnamefont {Gutmann}},\ and\ \bibinfo {author}
  {\bibfnamefont {D.~A.}\ \bibnamefont {Spielman}},\ }in\ \href
  {https://doi.org/10.1145/780542.780552} {\emph {\bibinfo {booktitle}
  {Proceedings of the Thirty-Fifth Annual ACM Symposium on Theory of
  Computing}}}\ (\bibinfo  {publisher} {Association for Computing Machinery},\
  \bibinfo {address} {New York, NY, USA},\ \bibinfo {year} {2003})\ p.\
  \bibinfo {pages} {59–68}\BibitemShut {NoStop}%
\bibitem [{\citenamefont {Childs}\ and\ \citenamefont
  {Goldstone}(2004)}]{Childs2004}%
  \BibitemOpen
  \bibfield  {author} {\bibinfo {author} {\bibfnamefont {A.~M.}\ \bibnamefont
  {Childs}}\ and\ \bibinfo {author} {\bibfnamefont {J.}~\bibnamefont
  {Goldstone}},\ }\href {https://doi.org/10.1103/PhysRevA.70.022314} {\bibfield
   {journal} {\bibinfo  {journal} {Phys. Rev. A}\ }\textbf {\bibinfo {volume}
  {70}},\ \bibinfo {pages} {022314} (\bibinfo {year} {2004})}\BibitemShut
  {NoStop}%
\bibitem [{\citenamefont {Childs}(2009)}]{Childs2009}%
  \BibitemOpen
  \bibfield  {author} {\bibinfo {author} {\bibfnamefont {A.~M.}\ \bibnamefont
  {Childs}},\ }\href {https://doi.org/10.1103/PhysRevLett.102.180501}
  {\bibfield  {journal} {\bibinfo  {journal} {Phys. Rev. Lett.}\ }\textbf
  {\bibinfo {volume} {102}},\ \bibinfo {pages} {180501} (\bibinfo {year}
  {2009})}\BibitemShut {NoStop}%
\bibitem [{\citenamefont {Venegas-Andraca}(2012)}]{Venegas2012}%
  \BibitemOpen
  \bibfield  {author} {\bibinfo {author} {\bibfnamefont {S.~E.}\ \bibnamefont
  {Venegas-Andraca}},\ }\href {https://doi.org/10.1007/s11128-012-0432-5}
  {\bibfield  {journal} {\bibinfo  {journal} {Quantum Information Processing}\
  }\textbf {\bibinfo {volume} {11}},\ \bibinfo {pages} {1015} (\bibinfo {year}
  {2012})}\BibitemShut {NoStop}%
\bibitem [{\citenamefont {Mark}\ \emph {et~al.}(2023)\citenamefont {Mark},
  \citenamefont {Choi}, \citenamefont {Shaw}, \citenamefont {Endres},\ and\
  \citenamefont {Choi}}]{Mark2023}%
  \BibitemOpen
  \bibfield  {author} {\bibinfo {author} {\bibfnamefont {D.~K.}\ \bibnamefont
  {Mark}}, \bibinfo {author} {\bibfnamefont {J.}~\bibnamefont {Choi}}, \bibinfo
  {author} {\bibfnamefont {A.~L.}\ \bibnamefont {Shaw}}, \bibinfo {author}
  {\bibfnamefont {M.}~\bibnamefont {Endres}},\ and\ \bibinfo {author}
  {\bibfnamefont {S.}~\bibnamefont {Choi}},\ }\href
  {https://doi.org/10.1103/PhysRevLett.131.110601} {\bibfield  {journal}
  {\bibinfo  {journal} {Phys. Rev. Lett.}\ }\textbf {\bibinfo {volume} {131}},\
  \bibinfo {pages} {110601} (\bibinfo {year} {2023})}\BibitemShut {NoStop}%
\bibitem [{\citenamefont {Preiss}\ \emph {et~al.}(2015)\citenamefont {Preiss},
  \citenamefont {Ma}, \citenamefont {Tai}, \citenamefont {Lukin}, \citenamefont
  {Rispoli}, \citenamefont {Zupancic}, \citenamefont {Lahini}, \citenamefont
  {Islam},\ and\ \citenamefont {Greiner}}]{Preiss-QWalk}%
  \BibitemOpen
  \bibfield  {author} {\bibinfo {author} {\bibfnamefont {P.~M.}\ \bibnamefont
  {Preiss}}, \bibinfo {author} {\bibfnamefont {R.}~\bibnamefont {Ma}}, \bibinfo
  {author} {\bibfnamefont {M.~E.}\ \bibnamefont {Tai}}, \bibinfo {author}
  {\bibfnamefont {A.}~\bibnamefont {Lukin}}, \bibinfo {author} {\bibfnamefont
  {M.}~\bibnamefont {Rispoli}}, \bibinfo {author} {\bibfnamefont
  {P.}~\bibnamefont {Zupancic}}, \bibinfo {author} {\bibfnamefont
  {Y.}~\bibnamefont {Lahini}}, \bibinfo {author} {\bibfnamefont
  {R.}~\bibnamefont {Islam}},\ and\ \bibinfo {author} {\bibfnamefont
  {M.}~\bibnamefont {Greiner}},\ }\href
  {https://doi.org/10.1126/science.1260364} {\bibfield  {journal} {\bibinfo
  {journal} {Science}\ }\textbf {\bibinfo {volume} {347}},\ \bibinfo {pages}
  {1229} (\bibinfo {year} {2015})}\BibitemShut {NoStop}%
\bibitem [{\citenamefont {Young}\ \emph {et~al.}(2022)\citenamefont {Young},
  \citenamefont {Eckner}, \citenamefont {Schine}, \citenamefont {Childs},\ and\
  \citenamefont {Kaufman}}]{Young2022}%
  \BibitemOpen
  \bibfield  {author} {\bibinfo {author} {\bibfnamefont {A.~W.}\ \bibnamefont
  {Young}}, \bibinfo {author} {\bibfnamefont {W.~J.}\ \bibnamefont {Eckner}},
  \bibinfo {author} {\bibfnamefont {N.}~\bibnamefont {Schine}}, \bibinfo
  {author} {\bibfnamefont {A.~M.}\ \bibnamefont {Childs}},\ and\ \bibinfo
  {author} {\bibfnamefont {A.~M.}\ \bibnamefont {Kaufman}},\ }\href
  {https://doi.org/10.1126/science.abo0608} {\bibfield  {journal} {\bibinfo
  {journal} {Science}\ }\textbf {\bibinfo {volume} {377}},\ \bibinfo {pages}
  {885} (\bibinfo {year} {2022})}\BibitemShut {NoStop}%
\bibitem [{\citenamefont {Gong}\ \emph {et~al.}(2021)\citenamefont {Gong},
  \citenamefont {Wang}, \citenamefont {Zha}, \citenamefont {Chen},
  \citenamefont {Huang}, \citenamefont {Wu}, \citenamefont {Zhu}, \citenamefont
  {Zhao}, \citenamefont {Li}, \citenamefont {Guo}, \citenamefont {Qian},
  \citenamefont {Ye}, \citenamefont {Chen}, \citenamefont {Ying}, \citenamefont
  {Yu}, \citenamefont {Fan}, \citenamefont {Wu}, \citenamefont {Su},
  \citenamefont {Deng}, \citenamefont {Rong}, \citenamefont {Zhang},
  \citenamefont {Cao}, \citenamefont {Lin}, \citenamefont {Xu}, \citenamefont
  {Sun}, \citenamefont {Guo}, \citenamefont {Li}, \citenamefont {Liang},
  \citenamefont {Bastidas}, \citenamefont {Nemoto}, \citenamefont {Munro},
  \citenamefont {Huo}, \citenamefont {Lu}, \citenamefont {Peng}, \citenamefont
  {Zhu},\ and\ \citenamefont {Pan}}]{Gong2021}%
  \BibitemOpen
  \bibfield  {author} {\bibinfo {author} {\bibfnamefont {M.}~\bibnamefont
  {Gong}}, \bibinfo {author} {\bibfnamefont {S.}~\bibnamefont {Wang}}, \bibinfo
  {author} {\bibfnamefont {C.}~\bibnamefont {Zha}}, \bibinfo {author}
  {\bibfnamefont {M.-C.}\ \bibnamefont {Chen}}, \bibinfo {author}
  {\bibfnamefont {H.-L.}\ \bibnamefont {Huang}}, \bibinfo {author}
  {\bibfnamefont {Y.}~\bibnamefont {Wu}}, \bibinfo {author} {\bibfnamefont
  {Q.}~\bibnamefont {Zhu}}, \bibinfo {author} {\bibfnamefont {Y.}~\bibnamefont
  {Zhao}}, \bibinfo {author} {\bibfnamefont {S.}~\bibnamefont {Li}}, \bibinfo
  {author} {\bibfnamefont {S.}~\bibnamefont {Guo}}, \bibinfo {author}
  {\bibfnamefont {H.}~\bibnamefont {Qian}}, \bibinfo {author} {\bibfnamefont
  {Y.}~\bibnamefont {Ye}}, \bibinfo {author} {\bibfnamefont {F.}~\bibnamefont
  {Chen}}, \bibinfo {author} {\bibfnamefont {C.}~\bibnamefont {Ying}}, \bibinfo
  {author} {\bibfnamefont {J.}~\bibnamefont {Yu}}, \bibinfo {author}
  {\bibfnamefont {D.}~\bibnamefont {Fan}}, \bibinfo {author} {\bibfnamefont
  {D.}~\bibnamefont {Wu}}, \bibinfo {author} {\bibfnamefont {H.}~\bibnamefont
  {Su}}, \bibinfo {author} {\bibfnamefont {H.}~\bibnamefont {Deng}}, \bibinfo
  {author} {\bibfnamefont {H.}~\bibnamefont {Rong}}, \bibinfo {author}
  {\bibfnamefont {K.}~\bibnamefont {Zhang}}, \bibinfo {author} {\bibfnamefont
  {S.}~\bibnamefont {Cao}}, \bibinfo {author} {\bibfnamefont {J.}~\bibnamefont
  {Lin}}, \bibinfo {author} {\bibfnamefont {Y.}~\bibnamefont {Xu}}, \bibinfo
  {author} {\bibfnamefont {L.}~\bibnamefont {Sun}}, \bibinfo {author}
  {\bibfnamefont {C.}~\bibnamefont {Guo}}, \bibinfo {author} {\bibfnamefont
  {N.}~\bibnamefont {Li}}, \bibinfo {author} {\bibfnamefont {F.}~\bibnamefont
  {Liang}}, \bibinfo {author} {\bibfnamefont {V.~M.}\ \bibnamefont {Bastidas}},
  \bibinfo {author} {\bibfnamefont {K.}~\bibnamefont {Nemoto}}, \bibinfo
  {author} {\bibfnamefont {W.~J.}\ \bibnamefont {Munro}}, \bibinfo {author}
  {\bibfnamefont {Y.-H.}\ \bibnamefont {Huo}}, \bibinfo {author} {\bibfnamefont
  {C.-Y.}\ \bibnamefont {Lu}}, \bibinfo {author} {\bibfnamefont {C.-Z.}\
  \bibnamefont {Peng}}, \bibinfo {author} {\bibfnamefont {X.}~\bibnamefont
  {Zhu}},\ and\ \bibinfo {author} {\bibfnamefont {J.-W.}\ \bibnamefont {Pan}},\
  }\href {https://doi.org/10.1126/science.abg7812} {\bibfield  {journal}
  {\bibinfo  {journal} {Science}\ }\textbf {\bibinfo {volume} {372}},\ \bibinfo
  {pages} {948} (\bibinfo {year} {2021})}\BibitemShut {NoStop}%
\bibitem [{\citenamefont {Schmitz}\ \emph {et~al.}(2009)\citenamefont
  {Schmitz}, \citenamefont {Matjeschk}, \citenamefont {Schneider},
  \citenamefont {Glueckert}, \citenamefont {Enderlein}, \citenamefont {Huber},\
  and\ \citenamefont {Schaetz}}]{Schmitz2009}%
  \BibitemOpen
  \bibfield  {author} {\bibinfo {author} {\bibfnamefont {H.}~\bibnamefont
  {Schmitz}}, \bibinfo {author} {\bibfnamefont {R.}~\bibnamefont {Matjeschk}},
  \bibinfo {author} {\bibfnamefont {C.}~\bibnamefont {Schneider}}, \bibinfo
  {author} {\bibfnamefont {J.}~\bibnamefont {Glueckert}}, \bibinfo {author}
  {\bibfnamefont {M.}~\bibnamefont {Enderlein}}, \bibinfo {author}
  {\bibfnamefont {T.}~\bibnamefont {Huber}},\ and\ \bibinfo {author}
  {\bibfnamefont {T.}~\bibnamefont {Schaetz}},\ }\href
  {https://doi.org/10.1103/PhysRevLett.103.090504} {\bibfield  {journal}
  {\bibinfo  {journal} {Phys. Rev. Lett.}\ }\textbf {\bibinfo {volume} {103}},\
  \bibinfo {pages} {090504} (\bibinfo {year} {2009})}\BibitemShut {NoStop}%
\bibitem [{\citenamefont {Z\"ahringer}\ \emph {et~al.}(2010)\citenamefont
  {Z\"ahringer}, \citenamefont {Kirchmair}, \citenamefont {Gerritsma},
  \citenamefont {Solano}, \citenamefont {Blatt},\ and\ \citenamefont
  {Roos}}]{Zahringer2010}%
  \BibitemOpen
  \bibfield  {author} {\bibinfo {author} {\bibfnamefont {F.}~\bibnamefont
  {Z\"ahringer}}, \bibinfo {author} {\bibfnamefont {G.}~\bibnamefont
  {Kirchmair}}, \bibinfo {author} {\bibfnamefont {R.}~\bibnamefont
  {Gerritsma}}, \bibinfo {author} {\bibfnamefont {E.}~\bibnamefont {Solano}},
  \bibinfo {author} {\bibfnamefont {R.}~\bibnamefont {Blatt}},\ and\ \bibinfo
  {author} {\bibfnamefont {C.~F.}\ \bibnamefont {Roos}},\ }\href
  {https://doi.org/10.1103/PhysRevLett.104.100503} {\bibfield  {journal}
  {\bibinfo  {journal} {Phys. Rev. Lett.}\ }\textbf {\bibinfo {volume} {104}},\
  \bibinfo {pages} {100503} (\bibinfo {year} {2010})}\BibitemShut {NoStop}%
\bibitem [{\citenamefont {Peruzzo}\ \emph {et~al.}(2010)\citenamefont
  {Peruzzo}, \citenamefont {Lobino}, \citenamefont {Matthews}, \citenamefont
  {Matsuda}, \citenamefont {Politi}, \citenamefont {Poulios}, \citenamefont
  {Zhou}, \citenamefont {Lahini}, \citenamefont {Ismail}, \citenamefont
  {Wörhoff}, \citenamefont {Bromberg}, \citenamefont {Silberberg},
  \citenamefont {Thompson},\ and\ \citenamefont {OBrien}}]{Peruzzo2010}%
  \BibitemOpen
  \bibfield  {author} {\bibinfo {author} {\bibfnamefont {A.}~\bibnamefont
  {Peruzzo}}, \bibinfo {author} {\bibfnamefont {M.}~\bibnamefont {Lobino}},
  \bibinfo {author} {\bibfnamefont {J.~C.~F.}\ \bibnamefont {Matthews}},
  \bibinfo {author} {\bibfnamefont {N.}~\bibnamefont {Matsuda}}, \bibinfo
  {author} {\bibfnamefont {A.}~\bibnamefont {Politi}}, \bibinfo {author}
  {\bibfnamefont {K.}~\bibnamefont {Poulios}}, \bibinfo {author} {\bibfnamefont
  {X.-Q.}\ \bibnamefont {Zhou}}, \bibinfo {author} {\bibfnamefont
  {Y.}~\bibnamefont {Lahini}}, \bibinfo {author} {\bibfnamefont
  {N.}~\bibnamefont {Ismail}}, \bibinfo {author} {\bibfnamefont
  {K.}~\bibnamefont {Wörhoff}}, \bibinfo {author} {\bibfnamefont
  {Y.}~\bibnamefont {Bromberg}}, \bibinfo {author} {\bibfnamefont
  {Y.}~\bibnamefont {Silberberg}}, \bibinfo {author} {\bibfnamefont {M.~G.}\
  \bibnamefont {Thompson}},\ and\ \bibinfo {author} {\bibfnamefont {J.~L.}\
  \bibnamefont {OBrien}},\ }\href {https://doi.org/10.1126/science.1193515}
  {\bibfield  {journal} {\bibinfo  {journal} {Science}\ }\textbf {\bibinfo
  {volume} {329}},\ \bibinfo {pages} {1500} (\bibinfo {year}
  {2010})}\BibitemShut {NoStop}%
\bibitem [{\citenamefont {Wang}\ and\ \citenamefont
  {Manouchehri}(2013)}]{Manouchehri2014}%
  \BibitemOpen
  \bibfield  {author} {\bibinfo {author} {\bibfnamefont {J.}~\bibnamefont
  {Wang}}\ and\ \bibinfo {author} {\bibfnamefont {K.}~\bibnamefont
  {Manouchehri}},\ }\href {https://doi.org/10.1007/978-3-642-36014-5}
  {\bibfield  {journal} {\bibinfo  {journal} {Heidelberg, Springer Berlin}\
  }\textbf {\bibinfo {volume} {10}},\ \bibinfo {pages} {978} (\bibinfo {year}
  {2013})}\BibitemShut {NoStop}%
\bibitem [{\citenamefont {Ben~Dahan}\ \emph {et~al.}(1996)\citenamefont
  {Ben~Dahan}, \citenamefont {Peik}, \citenamefont {Reichel}, \citenamefont
  {Castin},\ and\ \citenamefont {Salomon}}]{Dahan1996}%
  \BibitemOpen
  \bibfield  {author} {\bibinfo {author} {\bibfnamefont {M.}~\bibnamefont
  {Ben~Dahan}}, \bibinfo {author} {\bibfnamefont {E.}~\bibnamefont {Peik}},
  \bibinfo {author} {\bibfnamefont {J.}~\bibnamefont {Reichel}}, \bibinfo
  {author} {\bibfnamefont {Y.}~\bibnamefont {Castin}},\ and\ \bibinfo {author}
  {\bibfnamefont {C.}~\bibnamefont {Salomon}},\ }\href
  {https://doi.org/10.1103/PhysRevLett.76.4508} {\bibfield  {journal} {\bibinfo
   {journal} {Phys. Rev. Lett.}\ }\textbf {\bibinfo {volume} {76}},\ \bibinfo
  {pages} {4508} (\bibinfo {year} {1996})}\BibitemShut {NoStop}%
\bibitem [{\citenamefont {Geiger}\ \emph {et~al.}(2018)\citenamefont {Geiger},
  \citenamefont {Fujiwara}, \citenamefont {Singh}, \citenamefont {Senaratne},
  \citenamefont {Rajagopal}, \citenamefont {Lipatov}, \citenamefont
  {Shimasaki}, \citenamefont {Driben}, \citenamefont {Konotop}, \citenamefont
  {Meier},\ and\ \citenamefont {Weld}}]{Geiger2018}%
  \BibitemOpen
  \bibfield  {author} {\bibinfo {author} {\bibfnamefont {Z.~A.}\ \bibnamefont
  {Geiger}}, \bibinfo {author} {\bibfnamefont {K.~M.}\ \bibnamefont
  {Fujiwara}}, \bibinfo {author} {\bibfnamefont {K.}~\bibnamefont {Singh}},
  \bibinfo {author} {\bibfnamefont {R.}~\bibnamefont {Senaratne}}, \bibinfo
  {author} {\bibfnamefont {S.~V.}\ \bibnamefont {Rajagopal}}, \bibinfo {author}
  {\bibfnamefont {M.}~\bibnamefont {Lipatov}}, \bibinfo {author} {\bibfnamefont
  {T.}~\bibnamefont {Shimasaki}}, \bibinfo {author} {\bibfnamefont
  {R.}~\bibnamefont {Driben}}, \bibinfo {author} {\bibfnamefont {V.~V.}\
  \bibnamefont {Konotop}}, \bibinfo {author} {\bibfnamefont {T.}~\bibnamefont
  {Meier}},\ and\ \bibinfo {author} {\bibfnamefont {D.~M.}\ \bibnamefont
  {Weld}},\ }\href {https://doi.org/10.1103/PhysRevLett.120.213201} {\bibfield
  {journal} {\bibinfo  {journal} {Phys. Rev. Lett.}\ }\textbf {\bibinfo
  {volume} {120}},\ \bibinfo {pages} {213201} (\bibinfo {year}
  {2018})}\BibitemShut {NoStop}%
\bibitem [{\citenamefont {Mancini}\ \emph {et~al.}(2015)\citenamefont
  {Mancini}, \citenamefont {Pagano}, \citenamefont {Cappellini}, \citenamefont
  {Livi}, \citenamefont {Rider}, \citenamefont {Catani}, \citenamefont {Sias},
  \citenamefont {Zoller}, \citenamefont {Inguscio}, \citenamefont {Dalmonte},\
  and\ \citenamefont {Fallani}}]{Mancini2015}%
  \BibitemOpen
  \bibfield  {author} {\bibinfo {author} {\bibfnamefont {M.}~\bibnamefont
  {Mancini}}, \bibinfo {author} {\bibfnamefont {G.}~\bibnamefont {Pagano}},
  \bibinfo {author} {\bibfnamefont {G.}~\bibnamefont {Cappellini}}, \bibinfo
  {author} {\bibfnamefont {L.}~\bibnamefont {Livi}}, \bibinfo {author}
  {\bibfnamefont {M.}~\bibnamefont {Rider}}, \bibinfo {author} {\bibfnamefont
  {J.}~\bibnamefont {Catani}}, \bibinfo {author} {\bibfnamefont
  {C.}~\bibnamefont {Sias}}, \bibinfo {author} {\bibfnamefont {P.}~\bibnamefont
  {Zoller}}, \bibinfo {author} {\bibfnamefont {M.}~\bibnamefont {Inguscio}},
  \bibinfo {author} {\bibfnamefont {M.}~\bibnamefont {Dalmonte}},\ and\
  \bibinfo {author} {\bibfnamefont {L.}~\bibnamefont {Fallani}},\ }\href
  {https://doi.org/10.1126/science.aaa8736} {\bibfield  {journal} {\bibinfo
  {journal} {Science}\ }\textbf {\bibinfo {volume} {349}},\ \bibinfo {pages}
  {1510} (\bibinfo {year} {2015})}\BibitemShut {NoStop}%
\bibitem [{\citenamefont {An}\ \emph {et~al.}(2018)\citenamefont {An},
  \citenamefont {Meier}, \citenamefont {Ang'ong'a},\ and\ \citenamefont
  {Gadway}}]{an2018corr}%
  \BibitemOpen
  \bibfield  {author} {\bibinfo {author} {\bibfnamefont {F.~A.}\ \bibnamefont
  {An}}, \bibinfo {author} {\bibfnamefont {E.~J.}\ \bibnamefont {Meier}},
  \bibinfo {author} {\bibfnamefont {J.}~\bibnamefont {Ang'ong'a}},\ and\
  \bibinfo {author} {\bibfnamefont {B.}~\bibnamefont {Gadway}},\ }\href
  {https://doi.org/10.1103/PhysRevLett.120.040407} {\bibfield  {journal}
  {\bibinfo  {journal} {Phys. Rev. Lett.}\ }\textbf {\bibinfo {volume} {120}},\
  \bibinfo {pages} {040407} (\bibinfo {year} {2018})}\BibitemShut {NoStop}%
\bibitem [{\citenamefont {Yuan}\ \emph {et~al.}(2018)\citenamefont {Yuan},
  \citenamefont {Lin}, \citenamefont {Xiao},\ and\ \citenamefont
  {Fan}}]{Yuan:18}%
  \BibitemOpen
  \bibfield  {author} {\bibinfo {author} {\bibfnamefont {L.}~\bibnamefont
  {Yuan}}, \bibinfo {author} {\bibfnamefont {Q.}~\bibnamefont {Lin}}, \bibinfo
  {author} {\bibfnamefont {M.}~\bibnamefont {Xiao}},\ and\ \bibinfo {author}
  {\bibfnamefont {S.}~\bibnamefont {Fan}},\ }\href
  {https://doi.org/10.1364/OPTICA.5.001396} {\bibfield  {journal} {\bibinfo
  {journal} {Optica}\ }\textbf {\bibinfo {volume} {5}},\ \bibinfo {pages}
  {1396} (\bibinfo {year} {2018})}\BibitemShut {NoStop}%
\bibitem [{\citenamefont {Sundar}\ \emph {et~al.}(2019)\citenamefont {Sundar},
  \citenamefont {Thibodeau}, \citenamefont {Wang}, \citenamefont {Gadway},\
  and\ \citenamefont {Hazzard}}]{SundarPRA}%
  \BibitemOpen
  \bibfield  {author} {\bibinfo {author} {\bibfnamefont {B.}~\bibnamefont
  {Sundar}}, \bibinfo {author} {\bibfnamefont {M.}~\bibnamefont {Thibodeau}},
  \bibinfo {author} {\bibfnamefont {Z.}~\bibnamefont {Wang}}, \bibinfo {author}
  {\bibfnamefont {B.}~\bibnamefont {Gadway}},\ and\ \bibinfo {author}
  {\bibfnamefont {K.~R.~A.}\ \bibnamefont {Hazzard}},\ }\href
  {https://doi.org/10.1103/PhysRevA.99.013624} {\bibfield  {journal} {\bibinfo
  {journal} {Phys. Rev. A}\ }\textbf {\bibinfo {volume} {99}},\ \bibinfo
  {pages} {013624} (\bibinfo {year} {2019})}\BibitemShut {NoStop}%
\bibitem [{\citenamefont {Ozawa}\ and\ \citenamefont
  {Price}(2019)}]{Ozawa2019}%
  \BibitemOpen
  \bibfield  {author} {\bibinfo {author} {\bibfnamefont {T.}~\bibnamefont
  {Ozawa}}\ and\ \bibinfo {author} {\bibfnamefont {H.~M.}\ \bibnamefont
  {Price}},\ }\href {https://doi.org/10.1038/s42254-019-0045-3} {\bibfield
  {journal} {\bibinfo  {journal} {Nature Reviews Physics}\ }\textbf {\bibinfo
  {volume} {1}},\ \bibinfo {pages} {349} (\bibinfo {year} {2019})}\BibitemShut
  {NoStop}%
\bibitem [{\citenamefont {Kanungo}\ \emph {et~al.}(2022)\citenamefont
  {Kanungo}, \citenamefont {Whalen}, \citenamefont {Lu}, \citenamefont {Yuan},
  \citenamefont {Dasgupta}, \citenamefont {Dunning}, \citenamefont {Hazzard},\
  and\ \citenamefont {Killian}}]{Kanungo2022}%
  \BibitemOpen
  \bibfield  {author} {\bibinfo {author} {\bibfnamefont {S.~K.}\ \bibnamefont
  {Kanungo}}, \bibinfo {author} {\bibfnamefont {J.~D.}\ \bibnamefont {Whalen}},
  \bibinfo {author} {\bibfnamefont {Y.}~\bibnamefont {Lu}}, \bibinfo {author}
  {\bibfnamefont {M.}~\bibnamefont {Yuan}}, \bibinfo {author} {\bibfnamefont
  {S.}~\bibnamefont {Dasgupta}}, \bibinfo {author} {\bibfnamefont {F.~B.}\
  \bibnamefont {Dunning}}, \bibinfo {author} {\bibfnamefont {K.~R.~A.}\
  \bibnamefont {Hazzard}},\ and\ \bibinfo {author} {\bibfnamefont {T.~C.}\
  \bibnamefont {Killian}},\ }\href {https://doi.org/10.1038/s41467-022-28550-y}
  {\bibfield  {journal} {\bibinfo  {journal} {Nature Communications}\ }\textbf
  {\bibinfo {volume} {13}},\ \bibinfo {pages} {972} (\bibinfo {year}
  {2022})}\BibitemShut {NoStop}%
\bibitem [{\citenamefont {Oliver}\ \emph {et~al.}(2023)\citenamefont {Oliver},
  \citenamefont {Smith}, \citenamefont {Easton}, \citenamefont {Salerno},
  \citenamefont {Guarrera}, \citenamefont {Goldman}, \citenamefont
  {Barontini},\ and\ \citenamefont {Price}}]{Oliver2023}%
  \BibitemOpen
  \bibfield  {author} {\bibinfo {author} {\bibfnamefont {C.}~\bibnamefont
  {Oliver}}, \bibinfo {author} {\bibfnamefont {A.}~\bibnamefont {Smith}},
  \bibinfo {author} {\bibfnamefont {T.}~\bibnamefont {Easton}}, \bibinfo
  {author} {\bibfnamefont {G.}~\bibnamefont {Salerno}}, \bibinfo {author}
  {\bibfnamefont {V.}~\bibnamefont {Guarrera}}, \bibinfo {author}
  {\bibfnamefont {N.}~\bibnamefont {Goldman}}, \bibinfo {author} {\bibfnamefont
  {G.}~\bibnamefont {Barontini}},\ and\ \bibinfo {author} {\bibfnamefont
  {H.~M.}\ \bibnamefont {Price}},\ }\href
  {https://doi.org/10.1103/PhysRevResearch.5.033001} {\bibfield  {journal}
  {\bibinfo  {journal} {Phys. Rev. Res.}\ }\textbf {\bibinfo {volume} {5}},\
  \bibinfo {pages} {033001} (\bibinfo {year} {2023})}\BibitemShut {NoStop}%
\bibitem [{\citenamefont {Englebert}\ \emph {et~al.}(2023)\citenamefont
  {Englebert}, \citenamefont {Goldman}, \citenamefont {Erkintalo},
  \citenamefont {Mostaan}, \citenamefont {Gorza}, \citenamefont {Leo},\ and\
  \citenamefont {Fatome}}]{Englebert2023}%
  \BibitemOpen
  \bibfield  {author} {\bibinfo {author} {\bibfnamefont {N.}~\bibnamefont
  {Englebert}}, \bibinfo {author} {\bibfnamefont {N.}~\bibnamefont {Goldman}},
  \bibinfo {author} {\bibfnamefont {M.}~\bibnamefont {Erkintalo}}, \bibinfo
  {author} {\bibfnamefont {N.}~\bibnamefont {Mostaan}}, \bibinfo {author}
  {\bibfnamefont {S.-P.}\ \bibnamefont {Gorza}}, \bibinfo {author}
  {\bibfnamefont {F.}~\bibnamefont {Leo}},\ and\ \bibinfo {author}
  {\bibfnamefont {J.}~\bibnamefont {Fatome}},\ }\href
  {https://doi.org/https://doi.org/10.1038/s41567-023-02005-7} {\bibfield
  {journal} {\bibinfo  {journal} {Nature Physics}\ }\textbf {\bibinfo {volume}
  {19}},\ \bibinfo {pages} {1014} (\bibinfo {year} {2023})}\BibitemShut
  {NoStop}%
\bibitem [{\citenamefont {Deng}\ \emph {et~al.}(2022)\citenamefont {Deng},
  \citenamefont {Dong}, \citenamefont {Zhang}, \citenamefont {Wu},
  \citenamefont {Yuan}, \citenamefont {Zhu}, \citenamefont {Jin}, \citenamefont
  {Li}, \citenamefont {Wang}, \citenamefont {Cai}, \citenamefont {Song},
  \citenamefont {Wang}, \citenamefont {You},\ and\ \citenamefont
  {Wang}}]{Deng2022}%
  \BibitemOpen
  \bibfield  {author} {\bibinfo {author} {\bibfnamefont {J.}~\bibnamefont
  {Deng}}, \bibinfo {author} {\bibfnamefont {H.}~\bibnamefont {Dong}}, \bibinfo
  {author} {\bibfnamefont {C.}~\bibnamefont {Zhang}}, \bibinfo {author}
  {\bibfnamefont {Y.}~\bibnamefont {Wu}}, \bibinfo {author} {\bibfnamefont
  {J.}~\bibnamefont {Yuan}}, \bibinfo {author} {\bibfnamefont {X.}~\bibnamefont
  {Zhu}}, \bibinfo {author} {\bibfnamefont {F.}~\bibnamefont {Jin}}, \bibinfo
  {author} {\bibfnamefont {H.}~\bibnamefont {Li}}, \bibinfo {author}
  {\bibfnamefont {Z.}~\bibnamefont {Wang}}, \bibinfo {author} {\bibfnamefont
  {H.}~\bibnamefont {Cai}}, \bibinfo {author} {\bibfnamefont {C.}~\bibnamefont
  {Song}}, \bibinfo {author} {\bibfnamefont {H.}~\bibnamefont {Wang}}, \bibinfo
  {author} {\bibfnamefont {J.~Q.}\ \bibnamefont {You}},\ and\ \bibinfo {author}
  {\bibfnamefont {D.-W.}\ \bibnamefont {Wang}},\ }\href
  {https://doi.org/10.1126/science.ade6219} {\bibfield  {journal} {\bibinfo
  {journal} {Science}\ }\textbf {\bibinfo {volume} {378}},\ \bibinfo {pages}
  {966} (\bibinfo {year} {2022})}\BibitemShut {NoStop}%
\bibitem [{\citenamefont {Parto}\ \emph {et~al.}(2023)\citenamefont {Parto},
  \citenamefont {Leefmans}, \citenamefont {Williams}, \citenamefont {Nori},\
  and\ \citenamefont {Marandi}}]{Parto2023}%
  \BibitemOpen
  \bibfield  {author} {\bibinfo {author} {\bibfnamefont {M.}~\bibnamefont
  {Parto}}, \bibinfo {author} {\bibfnamefont {C.}~\bibnamefont {Leefmans}},
  \bibinfo {author} {\bibfnamefont {J.}~\bibnamefont {Williams}}, \bibinfo
  {author} {\bibfnamefont {F.}~\bibnamefont {Nori}},\ and\ \bibinfo {author}
  {\bibfnamefont {A.}~\bibnamefont {Marandi}},\ }\href
  {https://www.nature.com/articles/s41467-023-37065-z} {\bibfield  {journal}
  {\bibinfo  {journal} {Nature Communications}\ }\textbf {\bibinfo {volume}
  {14}},\ \bibinfo {pages} {1440} (\bibinfo {year} {2023})}\BibitemShut
  {NoStop}%
\bibitem [{\citenamefont {Chen}\ \emph
  {et~al.}(2024{\natexlab{a}})\citenamefont {Chen}, \citenamefont {Huang},
  \citenamefont {Velkovsky}, \citenamefont {Hazzard}, \citenamefont {Covey},\
  and\ \citenamefont {Gadway}}]{chen2024}%
  \BibitemOpen
  \bibfield  {author} {\bibinfo {author} {\bibfnamefont {T.}~\bibnamefont
  {Chen}}, \bibinfo {author} {\bibfnamefont {C.}~\bibnamefont {Huang}},
  \bibinfo {author} {\bibfnamefont {I.}~\bibnamefont {Velkovsky}}, \bibinfo
  {author} {\bibfnamefont {K.~R.~A.}\ \bibnamefont {Hazzard}}, \bibinfo
  {author} {\bibfnamefont {J.~P.}\ \bibnamefont {Covey}},\ and\ \bibinfo
  {author} {\bibfnamefont {B.}~\bibnamefont {Gadway}},\ }\href
  {https://doi.org/10.1038/s41467-024-46823-6} {\bibfield  {journal} {\bibinfo
  {journal} {Nature Communications}\ }\textbf {\bibinfo {volume} {15}},\
  \bibinfo {pages} {2675} (\bibinfo {year} {2024}{\natexlab{a}})}\BibitemShut
  {NoStop}%
\bibitem [{\citenamefont {Mueller}(2004)}]{Mueller-Escher}%
  \BibitemOpen
  \bibfield  {author} {\bibinfo {author} {\bibfnamefont {E.~J.}\ \bibnamefont
  {Mueller}},\ }\href {https://doi.org/10.1103/PhysRevA.70.041603} {\bibfield
  {journal} {\bibinfo  {journal} {Phys. Rev. A}\ }\textbf {\bibinfo {volume}
  {70}},\ \bibinfo {pages} {041603} (\bibinfo {year} {2004})}\BibitemShut
  {NoStop}%
\bibitem [{\citenamefont {Lu}\ \emph {et~al.}(2024)\citenamefont {Lu},
  \citenamefont {Wang}, \citenamefont {Kanungo}, \citenamefont {Yoshida},
  \citenamefont {Dunning},\ and\ \citenamefont {Killian}}]{Lu2024}%
  \BibitemOpen
  \bibfield  {author} {\bibinfo {author} {\bibfnamefont {Y.}~\bibnamefont
  {Lu}}, \bibinfo {author} {\bibfnamefont {C.}~\bibnamefont {Wang}}, \bibinfo
  {author} {\bibfnamefont {S.~K.}\ \bibnamefont {Kanungo}}, \bibinfo {author}
  {\bibfnamefont {S.}~\bibnamefont {Yoshida}}, \bibinfo {author} {\bibfnamefont
  {F.~B.}\ \bibnamefont {Dunning}},\ and\ \bibinfo {author} {\bibfnamefont
  {T.~C.}\ \bibnamefont {Killian}},\ }\href
  {https://doi.org/10.1103/PhysRevA.109.032801} {\bibfield  {journal} {\bibinfo
   {journal} {Phys. Rev. A}\ }\textbf {\bibinfo {volume} {109}},\ \bibinfo
  {pages} {032801} (\bibinfo {year} {2024})}\BibitemShut {NoStop}%
\bibitem [{\citenamefont {Chen}\ \emph
  {et~al.}(2024{\natexlab{b}})\citenamefont {Chen}, \citenamefont {Huang},
  \citenamefont {Velkovsky}, \citenamefont {Ozawa}, \citenamefont {Price},
  \citenamefont {Covey},\ and\ \citenamefont {Gadway}}]{chen2024caging}%
  \BibitemOpen
  \bibfield  {author} {\bibinfo {author} {\bibfnamefont {T.}~\bibnamefont
  {Chen}}, \bibinfo {author} {\bibfnamefont {C.}~\bibnamefont {Huang}},
  \bibinfo {author} {\bibfnamefont {I.}~\bibnamefont {Velkovsky}}, \bibinfo
  {author} {\bibfnamefont {T.}~\bibnamefont {Ozawa}}, \bibinfo {author}
  {\bibfnamefont {H.}~\bibnamefont {Price}}, \bibinfo {author} {\bibfnamefont
  {J.~P.}\ \bibnamefont {Covey}},\ and\ \bibinfo {author} {\bibfnamefont
  {B.}~\bibnamefont {Gadway}},\ }\href@noop {} {\bibinfo {title}
  {Interaction-driven breakdown of aharonov–bohm caging in flat-band rydberg
  lattices}} (\bibinfo {year} {2024}{\natexlab{b}}),\ \Eprint
  {https://arxiv.org/abs/2404.00737} {arXiv:2404.00737 [cond-mat.quant-gas]}
  \BibitemShut {NoStop}%
\bibitem [{\citenamefont {Mendez}\ \emph {et~al.}(1988)\citenamefont {Mendez},
  \citenamefont {Agull\'o-Rueda},\ and\ \citenamefont {Hong}}]{Mendez-stark}%
  \BibitemOpen
  \bibfield  {author} {\bibinfo {author} {\bibfnamefont {E.~E.}\ \bibnamefont
  {Mendez}}, \bibinfo {author} {\bibfnamefont {F.}~\bibnamefont
  {Agull\'o-Rueda}},\ and\ \bibinfo {author} {\bibfnamefont {J.~M.}\
  \bibnamefont {Hong}},\ }\href {https://doi.org/10.1103/PhysRevLett.60.2426}
  {\bibfield  {journal} {\bibinfo  {journal} {Phys. Rev. Lett.}\ }\textbf
  {\bibinfo {volume} {60}},\ \bibinfo {pages} {2426} (\bibinfo {year}
  {1988})}\BibitemShut {NoStop}%
\bibitem [{\citenamefont {Ang'ong'a}\ \emph {et~al.}(2022)\citenamefont
  {Ang'ong'a}, \citenamefont {Huang}, \citenamefont {Covey},\ and\
  \citenamefont {Gadway}}]{JacksonPRR}%
  \BibitemOpen
  \bibfield  {author} {\bibinfo {author} {\bibfnamefont {J.}~\bibnamefont
  {Ang'ong'a}}, \bibinfo {author} {\bibfnamefont {C.}~\bibnamefont {Huang}},
  \bibinfo {author} {\bibfnamefont {J.~P.}\ \bibnamefont {Covey}},\ and\
  \bibinfo {author} {\bibfnamefont {B.}~\bibnamefont {Gadway}},\ }\href
  {https://doi.org/10.1103/PhysRevResearch.4.013240} {\bibfield  {journal}
  {\bibinfo  {journal} {Phys. Rev. Res.}\ }\textbf {\bibinfo {volume} {4}},\
  \bibinfo {pages} {013240} (\bibinfo {year} {2022})}\BibitemShut {NoStop}%
\bibitem [{Sup()}]{SuppMats}%
  \BibitemOpen
  \href@noop {} {}\bibinfo {note} {See Supplementary Material at xxxxxx, which
  includes Refs.\cite{chen2024,Mueller-Escher}, for more experimental details
  and details on the theoretical formulation.}\BibitemShut {Stop}%
\bibitem [{\citenamefont {Price}\ \emph {et~al.}(2017)\citenamefont {Price},
  \citenamefont {Ozawa},\ and\ \citenamefont {Goldman}}]{Price-Shaking}%
  \BibitemOpen
  \bibfield  {author} {\bibinfo {author} {\bibfnamefont {H.~M.}\ \bibnamefont
  {Price}}, \bibinfo {author} {\bibfnamefont {T.}~\bibnamefont {Ozawa}},\ and\
  \bibinfo {author} {\bibfnamefont {N.}~\bibnamefont {Goldman}},\ }\href
  {https://doi.org/10.1103/PhysRevA.95.023607} {\bibfield  {journal} {\bibinfo
  {journal} {Phys. Rev. A}\ }\textbf {\bibinfo {volume} {95}},\ \bibinfo
  {pages} {023607} (\bibinfo {year} {2017})}\BibitemShut {NoStop}%
\bibitem [{\citenamefont {Thouless}\ \emph {et~al.}(1982)\citenamefont
  {Thouless}, \citenamefont {Kohmoto}, \citenamefont {Nightingale},\ and\
  \citenamefont {den Nijs}}]{Thouless-Twist}%
  \BibitemOpen
  \bibfield  {author} {\bibinfo {author} {\bibfnamefont {D.~J.}\ \bibnamefont
  {Thouless}}, \bibinfo {author} {\bibfnamefont {M.}~\bibnamefont {Kohmoto}},
  \bibinfo {author} {\bibfnamefont {M.~P.}\ \bibnamefont {Nightingale}},\ and\
  \bibinfo {author} {\bibfnamefont {M.}~\bibnamefont {den Nijs}},\ }\href
  {https://doi.org/10.1103/PhysRevLett.49.405} {\bibfield  {journal} {\bibinfo
  {journal} {Phys. Rev. Lett.}\ }\textbf {\bibinfo {volume} {49}},\ \bibinfo
  {pages} {405} (\bibinfo {year} {1982})}\BibitemShut {NoStop}%
\bibitem [{\citenamefont {Meinert}\ \emph {et~al.}(2016)\citenamefont
  {Meinert}, \citenamefont {Mark}, \citenamefont {Lauber}, \citenamefont
  {Daley},\ and\ \citenamefont {N\"agerl}}]{Meinert-Floquet}%
  \BibitemOpen
  \bibfield  {author} {\bibinfo {author} {\bibfnamefont {F.}~\bibnamefont
  {Meinert}}, \bibinfo {author} {\bibfnamefont {M.~J.}\ \bibnamefont {Mark}},
  \bibinfo {author} {\bibfnamefont {K.}~\bibnamefont {Lauber}}, \bibinfo
  {author} {\bibfnamefont {A.~J.}\ \bibnamefont {Daley}},\ and\ \bibinfo
  {author} {\bibfnamefont {H.-C.}\ \bibnamefont {N\"agerl}},\ }\href
  {https://doi.org/10.1103/PhysRevLett.116.205301} {\bibfield  {journal}
  {\bibinfo  {journal} {Phys. Rev. Lett.}\ }\textbf {\bibinfo {volume} {116}},\
  \bibinfo {pages} {205301} (\bibinfo {year} {2016})}\BibitemShut {NoStop}%
\bibitem [{\citenamefont {Winkler}\ \emph {et~al.}(2006)\citenamefont
  {Winkler}, \citenamefont {Thalhammer}, \citenamefont {Lang}, \citenamefont
  {Grimm}, \citenamefont {Hecker~Denschlag}, \citenamefont {Daley},
  \citenamefont {Kantian}, \citenamefont {B{\"u}chler},\ and\ \citenamefont
  {Zoller}}]{Winkler2006}%
  \BibitemOpen
  \bibfield  {author} {\bibinfo {author} {\bibfnamefont {K.}~\bibnamefont
  {Winkler}}, \bibinfo {author} {\bibfnamefont {G.}~\bibnamefont {Thalhammer}},
  \bibinfo {author} {\bibfnamefont {F.}~\bibnamefont {Lang}}, \bibinfo {author}
  {\bibfnamefont {R.}~\bibnamefont {Grimm}}, \bibinfo {author} {\bibfnamefont
  {J.}~\bibnamefont {Hecker~Denschlag}}, \bibinfo {author} {\bibfnamefont
  {A.~J.}\ \bibnamefont {Daley}}, \bibinfo {author} {\bibfnamefont
  {A.}~\bibnamefont {Kantian}}, \bibinfo {author} {\bibfnamefont {H.~P.}\
  \bibnamefont {B{\"u}chler}},\ and\ \bibinfo {author} {\bibfnamefont
  {P.}~\bibnamefont {Zoller}},\ }\href {https://doi.org/10.1038/nature04918}
  {\bibfield  {journal} {\bibinfo  {journal} {Nature}\ }\textbf {\bibinfo
  {volume} {441}},\ \bibinfo {pages} {853} (\bibinfo {year}
  {2006})}\BibitemShut {NoStop}%
\bibitem [{\citenamefont {Greschner}\ and\ \citenamefont
  {Santos}(2015)}]{anyon-hubb-theory}%
  \BibitemOpen
  \bibfield  {author} {\bibinfo {author} {\bibfnamefont {S.}~\bibnamefont
  {Greschner}}\ and\ \bibinfo {author} {\bibfnamefont {L.}~\bibnamefont
  {Santos}},\ }\href {https://doi.org/10.1103/PhysRevLett.115.053002}
  {\bibfield  {journal} {\bibinfo  {journal} {Phys. Rev. Lett.}\ }\textbf
  {\bibinfo {volume} {115}},\ \bibinfo {pages} {053002} (\bibinfo {year}
  {2015})}\BibitemShut {NoStop}%
\bibitem [{\citenamefont {Kwan}\ \emph {et~al.}(2023)\citenamefont {Kwan},
  \citenamefont {Segura}, \citenamefont {Li}, \citenamefont {Kim},
  \citenamefont {Gorshkov}, \citenamefont {Eckardt}, \citenamefont
  {Bakkali-Hassani},\ and\ \citenamefont {Greiner}}]{kwan2023realization}%
  \BibitemOpen
  \bibfield  {author} {\bibinfo {author} {\bibfnamefont {J.}~\bibnamefont
  {Kwan}}, \bibinfo {author} {\bibfnamefont {P.}~\bibnamefont {Segura}},
  \bibinfo {author} {\bibfnamefont {Y.}~\bibnamefont {Li}}, \bibinfo {author}
  {\bibfnamefont {S.}~\bibnamefont {Kim}}, \bibinfo {author} {\bibfnamefont
  {A.~V.}\ \bibnamefont {Gorshkov}}, \bibinfo {author} {\bibfnamefont
  {A.}~\bibnamefont {Eckardt}}, \bibinfo {author} {\bibfnamefont
  {B.}~\bibnamefont {Bakkali-Hassani}},\ and\ \bibinfo {author} {\bibfnamefont
  {M.}~\bibnamefont {Greiner}},\ }\href@noop {} {\bibinfo {title} {Realization
  of 1d anyons with arbitrary statistical phase}} (\bibinfo {year} {2023}),\
  \Eprint {https://arxiv.org/abs/2306.01737} {arXiv:2306.01737
  [cond-mat.quant-gas]} \BibitemShut {NoStop}%
\bibitem [{\citenamefont {Pan}\ and\ \citenamefont
  {Das~Sarma}(2023)}]{Kitaev-DasSarma}%
  \BibitemOpen
  \bibfield  {author} {\bibinfo {author} {\bibfnamefont {H.}~\bibnamefont
  {Pan}}\ and\ \bibinfo {author} {\bibfnamefont {S.}~\bibnamefont
  {Das~Sarma}},\ }\href {https://doi.org/10.1103/PhysRevB.107.035440}
  {\bibfield  {journal} {\bibinfo  {journal} {Phys. Rev. B}\ }\textbf {\bibinfo
  {volume} {107}},\ \bibinfo {pages} {035440} (\bibinfo {year}
  {2023})}\BibitemShut {NoStop}%
\bibitem [{\citenamefont {Zhao}\ \emph {et~al.}(2023)\citenamefont {Zhao},
  \citenamefont {Lee}, \citenamefont {Aliyu},\ and\ \citenamefont
  {Loh}}]{Huanqian-Floquet}%
  \BibitemOpen
  \bibfield  {author} {\bibinfo {author} {\bibfnamefont {L.}~\bibnamefont
  {Zhao}}, \bibinfo {author} {\bibfnamefont {M.~D.~K.}\ \bibnamefont {Lee}},
  \bibinfo {author} {\bibfnamefont {M.~M.}\ \bibnamefont {Aliyu}},\ and\
  \bibinfo {author} {\bibfnamefont {H.}~\bibnamefont {Loh}},\ }\href
  {https://doi.org/10.1038/s41467-023-42899-8} {\bibfield  {journal} {\bibinfo
  {journal} {Nature Communications}\ }\textbf {\bibinfo {volume} {14}},\
  \bibinfo {pages} {7128} (\bibinfo {year} {2023})}\BibitemShut {NoStop}%
\bibitem [{\citenamefont {Ates}\ \emph {et~al.}(2007)\citenamefont {Ates},
  \citenamefont {Pohl}, \citenamefont {Pattard},\ and\ \citenamefont
  {Rost}}]{Antiblockade}%
  \BibitemOpen
  \bibfield  {author} {\bibinfo {author} {\bibfnamefont {C.}~\bibnamefont
  {Ates}}, \bibinfo {author} {\bibfnamefont {T.}~\bibnamefont {Pohl}}, \bibinfo
  {author} {\bibfnamefont {T.}~\bibnamefont {Pattard}},\ and\ \bibinfo {author}
  {\bibfnamefont {J.~M.}\ \bibnamefont {Rost}},\ }\href
  {https://doi.org/10.1103/PhysRevLett.98.023002} {\bibfield  {journal}
  {\bibinfo  {journal} {Phys. Rev. Lett.}\ }\textbf {\bibinfo {volume} {98}},\
  \bibinfo {pages} {023002} (\bibinfo {year} {2007})}\BibitemShut {NoStop}%
\bibitem [{\citenamefont {Zhang}\ \emph {et~al.}(2023)\citenamefont {Zhang},
  \citenamefont {Kim}, \citenamefont {Mark}, \citenamefont {Choi},\ and\
  \citenamefont {Painter}}]{Zhang2023Painter}%
  \BibitemOpen
  \bibfield  {author} {\bibinfo {author} {\bibfnamefont {X.}~\bibnamefont
  {Zhang}}, \bibinfo {author} {\bibfnamefont {E.}~\bibnamefont {Kim}}, \bibinfo
  {author} {\bibfnamefont {D.~K.}\ \bibnamefont {Mark}}, \bibinfo {author}
  {\bibfnamefont {S.}~\bibnamefont {Choi}},\ and\ \bibinfo {author}
  {\bibfnamefont {O.}~\bibnamefont {Painter}},\ }\href
  {https://doi.org/10.1126/science.ade7651} {\bibfield  {journal} {\bibinfo
  {journal} {Science}\ }\textbf {\bibinfo {volume} {379}},\ \bibinfo {pages}
  {278} (\bibinfo {year} {2023})}\BibitemShut {NoStop}%
\bibitem [{\citenamefont {Young}\ \emph {et~al.}(2024)\citenamefont {Young},
  \citenamefont {Geller}, \citenamefont {Eckner}, \citenamefont {Schine},
  \citenamefont {Glancy}, \citenamefont {Knill},\ and\ \citenamefont
  {Kaufman}}]{Young2024sample}%
  \BibitemOpen
  \bibfield  {author} {\bibinfo {author} {\bibfnamefont {A.~W.}\ \bibnamefont
  {Young}}, \bibinfo {author} {\bibfnamefont {S.}~\bibnamefont {Geller}},
  \bibinfo {author} {\bibfnamefont {W.~J.}\ \bibnamefont {Eckner}}, \bibinfo
  {author} {\bibfnamefont {N.}~\bibnamefont {Schine}}, \bibinfo {author}
  {\bibfnamefont {S.}~\bibnamefont {Glancy}}, \bibinfo {author} {\bibfnamefont
  {E.}~\bibnamefont {Knill}},\ and\ \bibinfo {author} {\bibfnamefont {A.~M.}\
  \bibnamefont {Kaufman}},\ }\href
  {https://www.nature.com/articles/s41586-024-07304-4} {\bibfield  {journal}
  {\bibinfo  {journal} {Nature}\ }\textbf {\bibinfo {volume} {629}},\ \bibinfo
  {pages} {311} (\bibinfo {year} {2024})}\BibitemShut {NoStop}%
\bibitem [{\citenamefont {Zhong}\ \emph {et~al.}(2020)\citenamefont {Zhong},
  \citenamefont {Wang}, \citenamefont {Deng}, \citenamefont {Chen},
  \citenamefont {Peng}, \citenamefont {Luo}, \citenamefont {Qin}, \citenamefont
  {Wu}, \citenamefont {Ding}, \citenamefont {Hu}, \citenamefont {Hu},
  \citenamefont {Yang}, \citenamefont {Zhang}, \citenamefont {Li},
  \citenamefont {Li}, \citenamefont {Jiang}, \citenamefont {Gan}, \citenamefont
  {Yang}, \citenamefont {You}, \citenamefont {Wang}, \citenamefont {Li},
  \citenamefont {Liu}, \citenamefont {Lu},\ and\ \citenamefont
  {Pan}}]{Zhong2020sample}%
  \BibitemOpen
  \bibfield  {author} {\bibinfo {author} {\bibfnamefont {H.-S.}\ \bibnamefont
  {Zhong}}, \bibinfo {author} {\bibfnamefont {H.}~\bibnamefont {Wang}},
  \bibinfo {author} {\bibfnamefont {Y.-H.}\ \bibnamefont {Deng}}, \bibinfo
  {author} {\bibfnamefont {M.-C.}\ \bibnamefont {Chen}}, \bibinfo {author}
  {\bibfnamefont {L.-C.}\ \bibnamefont {Peng}}, \bibinfo {author}
  {\bibfnamefont {Y.-H.}\ \bibnamefont {Luo}}, \bibinfo {author} {\bibfnamefont
  {J.}~\bibnamefont {Qin}}, \bibinfo {author} {\bibfnamefont {D.}~\bibnamefont
  {Wu}}, \bibinfo {author} {\bibfnamefont {X.}~\bibnamefont {Ding}}, \bibinfo
  {author} {\bibfnamefont {Y.}~\bibnamefont {Hu}}, \bibinfo {author}
  {\bibfnamefont {P.}~\bibnamefont {Hu}}, \bibinfo {author} {\bibfnamefont
  {X.-Y.}\ \bibnamefont {Yang}}, \bibinfo {author} {\bibfnamefont {W.-J.}\
  \bibnamefont {Zhang}}, \bibinfo {author} {\bibfnamefont {H.}~\bibnamefont
  {Li}}, \bibinfo {author} {\bibfnamefont {Y.}~\bibnamefont {Li}}, \bibinfo
  {author} {\bibfnamefont {X.}~\bibnamefont {Jiang}}, \bibinfo {author}
  {\bibfnamefont {L.}~\bibnamefont {Gan}}, \bibinfo {author} {\bibfnamefont
  {G.}~\bibnamefont {Yang}}, \bibinfo {author} {\bibfnamefont {L.}~\bibnamefont
  {You}}, \bibinfo {author} {\bibfnamefont {Z.}~\bibnamefont {Wang}}, \bibinfo
  {author} {\bibfnamefont {L.}~\bibnamefont {Li}}, \bibinfo {author}
  {\bibfnamefont {N.-L.}\ \bibnamefont {Liu}}, \bibinfo {author} {\bibfnamefont
  {C.-Y.}\ \bibnamefont {Lu}},\ and\ \bibinfo {author} {\bibfnamefont {J.-W.}\
  \bibnamefont {Pan}},\ }\href {https://doi.org/10.1126/science.abe8770}
  {\bibfield  {journal} {\bibinfo  {journal} {Science}\ }\textbf {\bibinfo
  {volume} {370}},\ \bibinfo {pages} {1460} (\bibinfo {year}
  {2020})}\BibitemShut {NoStop}%
\bibitem [{\citenamefont {Barredo}\ \emph {et~al.}(2020)\citenamefont
  {Barredo}, \citenamefont {Lienhard}, \citenamefont {Scholl}, \citenamefont
  {de~L\'es\'eleuc}, \citenamefont {Boulier}, \citenamefont {Browaeys},\ and\
  \citenamefont {Lahaye}}]{Barredo2020}%
  \BibitemOpen
  \bibfield  {author} {\bibinfo {author} {\bibfnamefont {D.}~\bibnamefont
  {Barredo}}, \bibinfo {author} {\bibfnamefont {V.}~\bibnamefont {Lienhard}},
  \bibinfo {author} {\bibfnamefont {P.}~\bibnamefont {Scholl}}, \bibinfo
  {author} {\bibfnamefont {S.}~\bibnamefont {de~L\'es\'eleuc}}, \bibinfo
  {author} {\bibfnamefont {T.}~\bibnamefont {Boulier}}, \bibinfo {author}
  {\bibfnamefont {A.}~\bibnamefont {Browaeys}},\ and\ \bibinfo {author}
  {\bibfnamefont {T.}~\bibnamefont {Lahaye}},\ }\href
  {https://doi.org/10.1103/PhysRevLett.124.023201} {\bibfield  {journal}
  {\bibinfo  {journal} {Phys. Rev. Lett.}\ }\textbf {\bibinfo {volume} {124}},\
  \bibinfo {pages} {023201} (\bibinfo {year} {2020})}\BibitemShut {NoStop}%
\bibitem [{\citenamefont {Wilson}\ \emph {et~al.}(2022)\citenamefont {Wilson},
  \citenamefont {Saskin}, \citenamefont {Meng}, \citenamefont {Ma},
  \citenamefont {Dilip}, \citenamefont {Burgers},\ and\ \citenamefont
  {Thompson}}]{Wilson2022}%
  \BibitemOpen
  \bibfield  {author} {\bibinfo {author} {\bibfnamefont {J.~T.}\ \bibnamefont
  {Wilson}}, \bibinfo {author} {\bibfnamefont {S.}~\bibnamefont {Saskin}},
  \bibinfo {author} {\bibfnamefont {Y.}~\bibnamefont {Meng}}, \bibinfo {author}
  {\bibfnamefont {S.}~\bibnamefont {Ma}}, \bibinfo {author} {\bibfnamefont
  {R.}~\bibnamefont {Dilip}}, \bibinfo {author} {\bibfnamefont {A.~P.}\
  \bibnamefont {Burgers}},\ and\ \bibinfo {author} {\bibfnamefont {J.~D.}\
  \bibnamefont {Thompson}},\ }\href
  {https://doi.org/10.1103/PhysRevLett.128.033201} {\bibfield  {journal}
  {\bibinfo  {journal} {Phys. Rev. Lett.}\ }\textbf {\bibinfo {volume} {128}},\
  \bibinfo {pages} {033201} (\bibinfo {year} {2022})}\BibitemShut {NoStop}%
\bibitem [{\citenamefont {Thompson}\ \emph {et~al.}(2013)\citenamefont
  {Thompson}, \citenamefont {Tiecke}, \citenamefont {Zibrov}, \citenamefont
  {Vuletic},\ and\ \citenamefont {Lukin}}]{Thompson13}%
  \BibitemOpen
  \bibfield  {author} {\bibinfo {author} {\bibfnamefont {J.~D.}\ \bibnamefont
  {Thompson}}, \bibinfo {author} {\bibfnamefont {T.~G.}\ \bibnamefont
  {Tiecke}}, \bibinfo {author} {\bibfnamefont {A.~S.}\ \bibnamefont {Zibrov}},
  \bibinfo {author} {\bibfnamefont {V.}~\bibnamefont {Vuletic}},\ and\ \bibinfo
  {author} {\bibfnamefont {M.~D.}\ \bibnamefont {Lukin}},\ }\href
  {https://doi.org/10.1103/PhysRevLett.110.133001} {\bibfield  {journal}
  {\bibinfo  {journal} {Phys. Rev. Lett.}\ }\textbf {\bibinfo {volume} {110}},\
  \bibinfo {pages} {133001} (\bibinfo {year} {2013})}\BibitemShut {NoStop}%
\bibitem [{\citenamefont {Kaufman}\ \emph {et~al.}(2012)\citenamefont
  {Kaufman}, \citenamefont {Lester},\ and\ \citenamefont {Regal}}]{kaufman12}%
  \BibitemOpen
  \bibfield  {author} {\bibinfo {author} {\bibfnamefont {A.~M.}\ \bibnamefont
  {Kaufman}}, \bibinfo {author} {\bibfnamefont {B.~J.}\ \bibnamefont
  {Lester}},\ and\ \bibinfo {author} {\bibfnamefont {C.~A.}\ \bibnamefont
  {Regal}},\ }\href {https://doi.org/10.1103/PhysRevX.2.041014} {\bibfield
  {journal} {\bibinfo  {journal} {Phys. Rev. X}\ }\textbf {\bibinfo {volume}
  {2}},\ \bibinfo {pages} {041014} (\bibinfo {year} {2012})}\BibitemShut
  {NoStop}%
\bibitem [{\citenamefont {Wu}\ \emph {et~al.}(2023)\citenamefont {Wu},
  \citenamefont {Richaud}, \citenamefont {Raimond}, \citenamefont {Brune},\
  and\ \citenamefont {Gleyzes}}]{Wu2023}%
  \BibitemOpen
  \bibfield  {author} {\bibinfo {author} {\bibfnamefont {H.}~\bibnamefont
  {Wu}}, \bibinfo {author} {\bibfnamefont {R.}~\bibnamefont {Richaud}},
  \bibinfo {author} {\bibfnamefont {J.-M.}\ \bibnamefont {Raimond}}, \bibinfo
  {author} {\bibfnamefont {M.}~\bibnamefont {Brune}},\ and\ \bibinfo {author}
  {\bibfnamefont {S.}~\bibnamefont {Gleyzes}},\ }\href
  {https://doi.org/10.1103/PhysRevLett.130.023202} {\bibfield  {journal}
  {\bibinfo  {journal} {Phys. Rev. Lett.}\ }\textbf {\bibinfo {volume} {130}},\
  \bibinfo {pages} {023202} (\bibinfo {year} {2023})}\BibitemShut {NoStop}%
\bibitem [{\citenamefont {Schymik}\ \emph {et~al.}(2021)\citenamefont
  {Schymik}, \citenamefont {Pancaldi}, \citenamefont {Nogrette}, \citenamefont
  {Barredo}, \citenamefont {Paris}, \citenamefont {Browaeys},\ and\
  \citenamefont {Lahaye}}]{Schymik2021}%
  \BibitemOpen
  \bibfield  {author} {\bibinfo {author} {\bibfnamefont {K.-N.}\ \bibnamefont
  {Schymik}}, \bibinfo {author} {\bibfnamefont {S.}~\bibnamefont {Pancaldi}},
  \bibinfo {author} {\bibfnamefont {F.}~\bibnamefont {Nogrette}}, \bibinfo
  {author} {\bibfnamefont {D.}~\bibnamefont {Barredo}}, \bibinfo {author}
  {\bibfnamefont {J.}~\bibnamefont {Paris}}, \bibinfo {author} {\bibfnamefont
  {A.}~\bibnamefont {Browaeys}},\ and\ \bibinfo {author} {\bibfnamefont
  {T.}~\bibnamefont {Lahaye}},\ }\href
  {https://doi.org/10.1103/PhysRevApplied.16.034013} {\bibfield  {journal}
  {\bibinfo  {journal} {Phys. Rev. Appl.}\ }\textbf {\bibinfo {volume} {16}},\
  \bibinfo {pages} {034013} (\bibinfo {year} {2021})}\BibitemShut {NoStop}%
\bibitem [{\citenamefont {Abanin}\ \emph {et~al.}(2019)\citenamefont {Abanin},
  \citenamefont {Altman}, \citenamefont {Bloch},\ and\ \citenamefont
  {Serbyn}}]{Abanin2019}%
  \BibitemOpen
  \bibfield  {author} {\bibinfo {author} {\bibfnamefont {D.~A.}\ \bibnamefont
  {Abanin}}, \bibinfo {author} {\bibfnamefont {E.}~\bibnamefont {Altman}},
  \bibinfo {author} {\bibfnamefont {I.}~\bibnamefont {Bloch}},\ and\ \bibinfo
  {author} {\bibfnamefont {M.}~\bibnamefont {Serbyn}},\ }\href
  {https://doi.org/10.1103/RevModPhys.91.021001} {\bibfield  {journal}
  {\bibinfo  {journal} {Rev. Mod. Phys.}\ }\textbf {\bibinfo {volume} {91}},\
  \bibinfo {pages} {021001} (\bibinfo {year} {2019})}\BibitemShut {NoStop}%
\bibitem [{\citenamefont {Schreiber}\ \emph {et~al.}(2015)\citenamefont
  {Schreiber}, \citenamefont {Hodgman}, \citenamefont {Bordia}, \citenamefont
  {Lüschen}, \citenamefont {Fischer}, \citenamefont {Vosk}, \citenamefont
  {Altman}, \citenamefont {Schneider},\ and\ \citenamefont
  {Bloch}}]{Schreiber2015}%
  \BibitemOpen
  \bibfield  {author} {\bibinfo {author} {\bibfnamefont {M.}~\bibnamefont
  {Schreiber}}, \bibinfo {author} {\bibfnamefont {S.~S.}\ \bibnamefont
  {Hodgman}}, \bibinfo {author} {\bibfnamefont {P.}~\bibnamefont {Bordia}},
  \bibinfo {author} {\bibfnamefont {H.~P.}\ \bibnamefont {Lüschen}}, \bibinfo
  {author} {\bibfnamefont {M.~H.}\ \bibnamefont {Fischer}}, \bibinfo {author}
  {\bibfnamefont {R.}~\bibnamefont {Vosk}}, \bibinfo {author} {\bibfnamefont
  {E.}~\bibnamefont {Altman}}, \bibinfo {author} {\bibfnamefont
  {U.}~\bibnamefont {Schneider}},\ and\ \bibinfo {author} {\bibfnamefont
  {I.}~\bibnamefont {Bloch}},\ }\href {https://doi.org/10.1126/science.aaa7432}
  {\bibfield  {journal} {\bibinfo  {journal} {Science}\ }\textbf {\bibinfo
  {volume} {349}},\ \bibinfo {pages} {842} (\bibinfo {year}
  {2015})}\BibitemShut {NoStop}%
\bibitem [{\citenamefont {Lukin}\ \emph {et~al.}(2019)\citenamefont {Lukin},
  \citenamefont {Rispoli}, \citenamefont {Schittko}, \citenamefont {Tai},
  \citenamefont {Kaufman}, \citenamefont {Choi}, \citenamefont {Khemani},
  \citenamefont {Léonard},\ and\ \citenamefont {Greiner}}]{Lukin2019}%
  \BibitemOpen
  \bibfield  {author} {\bibinfo {author} {\bibfnamefont {A.}~\bibnamefont
  {Lukin}}, \bibinfo {author} {\bibfnamefont {M.}~\bibnamefont {Rispoli}},
  \bibinfo {author} {\bibfnamefont {R.}~\bibnamefont {Schittko}}, \bibinfo
  {author} {\bibfnamefont {M.~E.}\ \bibnamefont {Tai}}, \bibinfo {author}
  {\bibfnamefont {A.~M.}\ \bibnamefont {Kaufman}}, \bibinfo {author}
  {\bibfnamefont {S.}~\bibnamefont {Choi}}, \bibinfo {author} {\bibfnamefont
  {V.}~\bibnamefont {Khemani}}, \bibinfo {author} {\bibfnamefont
  {J.}~\bibnamefont {Léonard}},\ and\ \bibinfo {author} {\bibfnamefont
  {M.}~\bibnamefont {Greiner}},\ }\href
  {https://doi.org/10.1126/science.aau0818} {\bibfield  {journal} {\bibinfo
  {journal} {Science}\ }\textbf {\bibinfo {volume} {364}},\ \bibinfo {pages}
  {256} (\bibinfo {year} {2019})}\BibitemShut {NoStop}%
\bibitem [{\citenamefont {Yao}\ \emph {et~al.}(2016)\citenamefont {Yao},
  \citenamefont {Laumann}, \citenamefont {Cirac}, \citenamefont {Lukin},\ and\
  \citenamefont {Moore}}]{Yao2016}%
  \BibitemOpen
  \bibfield  {author} {\bibinfo {author} {\bibfnamefont {N.~Y.}\ \bibnamefont
  {Yao}}, \bibinfo {author} {\bibfnamefont {C.~R.}\ \bibnamefont {Laumann}},
  \bibinfo {author} {\bibfnamefont {J.~I.}\ \bibnamefont {Cirac}}, \bibinfo
  {author} {\bibfnamefont {M.~D.}\ \bibnamefont {Lukin}},\ and\ \bibinfo
  {author} {\bibfnamefont {J.~E.}\ \bibnamefont {Moore}},\ }\href
  {https://doi.org/10.1103/PhysRevLett.117.240601} {\bibfield  {journal}
  {\bibinfo  {journal} {Phys. Rev. Lett.}\ }\textbf {\bibinfo {volume} {117}},\
  \bibinfo {pages} {240601} (\bibinfo {year} {2016})}\BibitemShut {NoStop}%
\bibitem [{\citenamefont {Morong}\ \emph {et~al.}(2021)\citenamefont {Morong},
  \citenamefont {Liu}, \citenamefont {Becker}, \citenamefont {Collins},
  \citenamefont {Feng}, \citenamefont {Kyprianidis}, \citenamefont {Pagano},
  \citenamefont {You}, \citenamefont {Gorshkov},\ and\ \citenamefont
  {Monroe}}]{Morong2021}%
  \BibitemOpen
  \bibfield  {author} {\bibinfo {author} {\bibfnamefont {W.}~\bibnamefont
  {Morong}}, \bibinfo {author} {\bibfnamefont {F.}~\bibnamefont {Liu}},
  \bibinfo {author} {\bibfnamefont {P.}~\bibnamefont {Becker}}, \bibinfo
  {author} {\bibfnamefont {K.~S.}\ \bibnamefont {Collins}}, \bibinfo {author}
  {\bibfnamefont {L.}~\bibnamefont {Feng}}, \bibinfo {author} {\bibfnamefont
  {A.}~\bibnamefont {Kyprianidis}}, \bibinfo {author} {\bibfnamefont
  {G.}~\bibnamefont {Pagano}}, \bibinfo {author} {\bibfnamefont
  {T.}~\bibnamefont {You}}, \bibinfo {author} {\bibfnamefont {A.~V.}\
  \bibnamefont {Gorshkov}},\ and\ \bibinfo {author} {\bibfnamefont
  {C.}~\bibnamefont {Monroe}},\ }\href
  {https://doi.org/10.1038/s41586-021-03988-0} {\bibfield  {journal} {\bibinfo
  {journal} {Nature}\ }\textbf {\bibinfo {volume} {599}},\ \bibinfo {pages}
  {393} (\bibinfo {year} {2021})}\BibitemShut {NoStop}%
\bibitem [{\citenamefont {Rachel}(2018)}]{Rachel2018}%
  \BibitemOpen
  \bibfield  {author} {\bibinfo {author} {\bibfnamefont {S.}~\bibnamefont
  {Rachel}},\ }\href {https://doi.org/10.1088/1361-6633/aad6a6} {\bibfield
  {journal} {\bibinfo  {journal} {Reports on Progress in Physics}\ }\textbf
  {\bibinfo {volume} {81}},\ \bibinfo {pages} {116501} (\bibinfo {year}
  {2018})}\BibitemShut {NoStop}%
\bibitem [{\citenamefont {Walter}\ \emph {et~al.}(2023)\citenamefont {Walter},
  \citenamefont {Zhu}, \citenamefont {G{\"a}chter}, \citenamefont {Minguzzi},
  \citenamefont {Roschinski}, \citenamefont {Sandholzer}, \citenamefont
  {Viebahn},\ and\ \citenamefont {Esslinger}}]{Walter2023}%
  \BibitemOpen
  \bibfield  {author} {\bibinfo {author} {\bibfnamefont {A.-S.}\ \bibnamefont
  {Walter}}, \bibinfo {author} {\bibfnamefont {Z.}~\bibnamefont {Zhu}},
  \bibinfo {author} {\bibfnamefont {M.}~\bibnamefont {G{\"a}chter}}, \bibinfo
  {author} {\bibfnamefont {J.}~\bibnamefont {Minguzzi}}, \bibinfo {author}
  {\bibfnamefont {S.}~\bibnamefont {Roschinski}}, \bibinfo {author}
  {\bibfnamefont {K.}~\bibnamefont {Sandholzer}}, \bibinfo {author}
  {\bibfnamefont {K.}~\bibnamefont {Viebahn}},\ and\ \bibinfo {author}
  {\bibfnamefont {T.}~\bibnamefont {Esslinger}},\ }\href
  {https://doi.org/10.1038/s41567-023-02145-w} {\bibfield  {journal} {\bibinfo
  {journal} {Nature Physics}\ }\textbf {\bibinfo {volume} {19}},\ \bibinfo
  {pages} {1471} (\bibinfo {year} {2023})}\BibitemShut {NoStop}%
\end{thebibliography}%

\renewcommand{\thesection}{\Alph{section}}
\renewcommand{\thefigure}{S\arabic{figure}}
\renewcommand{\thetable}{S\Roman{table}}
\setcounter{figure}{0}
\renewcommand{\theequation}{S\arabic{equation}}
\renewcommand{\thepage}{S\arabic{page}}
\setcounter{equation}{0}
\setcounter{page}{1}

\clearpage

\begin{widetext}
\appendix

\section{\large Supplemental Material for ``Quantum walks and \\ correlated dynamics in an interacting synthetic Rydberg lattice"} 

\vspace{5mm}

\section{Calibration of the global flux for staircase lattice with $\Delta=0$}

The population dynamics strongly depend on the global flux in the 8-state ring structure with $\Delta=0$ in Fig.~2(a). Here we focus on the zero flux case. In the experiment, we adjust the phase $\phi_{01}$ of the microwave frequency tone that drives $\ket{0}\leftrightarrow\ket{1}$ transition (relative to the other tones, which all start with zero phase at the source) to achieve zero global flux. The calibration of the global flux and its dependence on $\phi_{01}$ is based on the measured population dynamics of single atoms. Figure~\ref{FIG:figs5} shows the population in state $\ket{4}$ after an evolution time of $2.25~\mu{\rm s}$ ($\sim 2h/\Omega$ with $\Omega/h=0.90(2)~{\rm MHz}$) for different $\phi_{01}$. Since we expect $P_4$ to peak for zero global flux at $t=2h/\Omega$, we use $\phi_{01}=0.60(2)\pi$ (based on the simple Gaussian fit shown in Fig.~\ref{FIG:figs5}) to achieve zero flux for all the measurements shown in Fig.~2 of the main text. 

\begin{figure*}[h]	
    \includegraphics[width=0.5\textwidth]{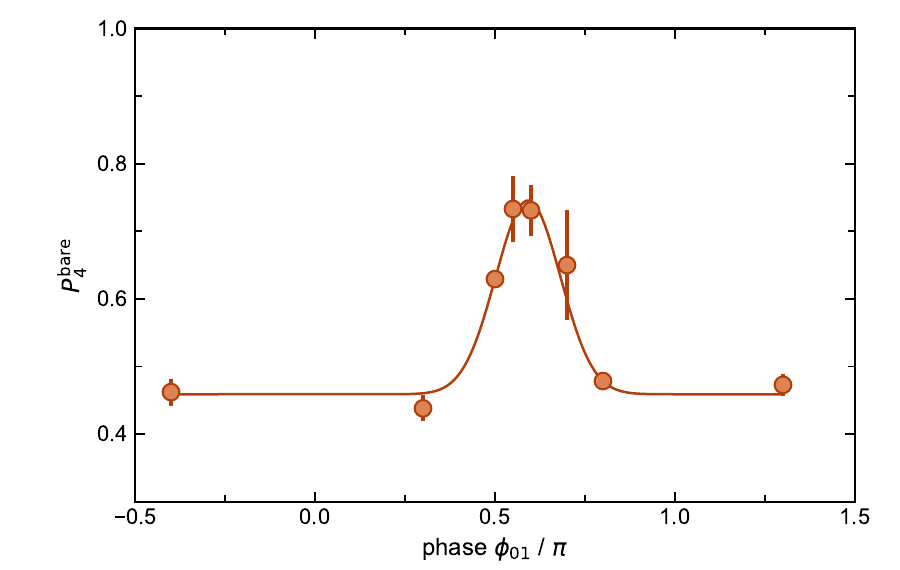}
	\caption{\textbf{Calibration of the global flux for 8-state ring structure.}
        Measured population in state $\ket{4}$, $P_4$, after an evolution time of $2.25~\mu{\rm s}$ under different values of the phase $\phi_{01}$. The solid line indicates the Gaussian fit to the data points. The error bars represent the standard errors from several independent measurements.
}
\label{FIG:figs5}
\end{figure*}

\section{Time-dependent Hamiltonian for 8-state Escher's staircase}

For the 8-state ring structure in Fig.~2 in the main text with nonzero tilting, the Hamiltonian is still time-dependent after applying a proper unitary transformation \cite{Mueller-Escher}
\begin{equation}
  H_{\rm pbc} = \sum_{j\in[-3,3]} \frac{\Omega}{2}\left(c^\dagger_j c_{j+1} + {\rm h.c.}\right) + \sum_j j\Delta c^\dagger_j c_j  + \left(\frac{\Omega}{2}e^{-i8\Delta t/\hbar}c^\dagger_4 c_{-3} + {\rm h.c.}\right)
\end{equation}
where $\Omega$ is the Rabi coupling strength and $\Delta$ the energy tilting.

\section{Details of the interaction terms}

As discussed in \cite{chen2024}, two different dipolar exchange processes should be addressed in our setup: (i) resonant state-conserved flip-flop interactions for $\ket{i}_A\ket{j}_B \leftrightarrow \ket{j}_A\ket{i}_B$, (ii) non-resonant state-changing interactions for $\ket{i}_A\ket{j'}_B \leftrightarrow \ket{j}_A\ket{i'}_B$ with a detuning $\Delta_{ij}^{i'j'}$. For the states employed to implement our synthetic lattice in Fig.~1(a), the smallest detuning $|\Delta_{ij}^{i'j'}|_{\rm min} \sim h\times 50~{\rm MHz}$ under the quantization $B$-field of $\sim 27~{\rm G}$. Since we work in the weak interaction regime with the maximum interaction strength $V=h\times2.7~{\rm MHz}$, i.e., $V \ll |\Delta_{ij}^{i'j'}|_{\rm min}$, the non-resonant state-changing interaction terms have negligible effect on the pair dynamics and consequently can be safely excluded. In our numerical simulations, we directly use the flip-flop interacting Hamiltonian (2) in the main text. The interaction strength $V_{ij} \propto C_3^{ij}/d^3_{AB}$ with $d_{AB}$ the spatial separation of the atom pair. Similar to our previous work \cite{chen2024}, we scale all $V_{ij}$ to the calibrated value $V=V_{0,-1}$ according to the relevant calculated $C_3^{ij}$ coefficients, as listed in Table \ref{tabs1}. 

\begin{table}[h]
\begin{tabular}{l|c||l|c}
 \hline
 \hline
 $\ket{i}\ket{j}$ & $C_3$ & $\ket{i}\ket{j}$ & $C_3$ \\[0.25em]\hline
 $\ket{0}\ket{1}$ & -756.4 &  & \\[0.25em]
 $\ket{0}\ket{-1}$ & 756.4 &  & \\[0.25em]\hline
 $\ket{1}\ket{2}$ & -639.1 & $\ket{-1}\ket{2}$ & 639.1 \\[0.25em]
 $\ket{1}\ket{-2}$ & 639.1 & $\ket{-1}\ket{-2}$ & -639.1 \\[0.25em]\hline
 $\ket{2}\ket{3}$ & -834.4 & $\ket{-2}\ket{3}$ & 834.4 \\[0.25em]
 $\ket{2}\ket{-3}$ & 834.4 & $\ket{-2}\ket{-3}$ & -834.4 \\[0.25em]\hline
 $\ket{3}\ket{4}$ & -705.0 & $\ket{-3}\ket{4}$ & 705.0 \\[0.25em]
 $\ket{3}\ket{-4}$ & 705.0 & $\ket{-3}\ket{-4}$ & -705.0 \\[0.25em]\hline
 \hline
\end{tabular}
\caption{\textbf{Calculated $C_3$ coefficients (units of $\rm{MHz} \ \mu{\rm m}^3$) for the resonant dipolar exchange interaction terms.} 
\label{tabs1}
}
\end{table}

\section{Renormalization of the experimental measurements}

As discussed in Ref.~\cite{chen2024}, the primary data we measure for the state population dynamics has a lower contrast as compared to the renormalized data presented in the main text. There are two main effects that reduce the contrast of the raw population dynamics data.
First, the data typically features an average upper ``ceiling'' value $P_u$, which stems from inefficiency of STIRAP, as well as loss during release-and-recapture. There is also a lower baseline of the measurements, having an average value $P_l$, that we believe stems from the decay (and subsequent recapture) of the short-lived Rydberg states, which results in the non-depumped Rydberg states having some probability to appear bright to subsequent fluorescence detection. These infidelities limit the contrast of state population dynamics.

For the averaged population dynamics in non-interacting singles ($P_i=\langle c_i^\dagger c_i\rangle$) and the interacting pairs [$P_i = \frac{1}{2}(\langle c_{i,A}^\dagger c_{i,A}\otimes I_B\rangle + \langle I_A\otimes c_{i,B}^\dagger c_{i,B}\rangle)$], we renormalize the measured $P_i^{\rm bare}$ to $P_i = (P_i^{\rm bare}-P_l)/(P_u-P_l)$ with $P_u = 0.93(1)$ and $P_l = 0.32(1)$. For the pair state dynamics, i.e., $P_{\ket{0,0}} = \langle c_{0,A}^\dagger c_{0,A}\otimes c_{0,B}^\dagger c_{0,B}\rangle$, we renormalize with $P_u = 0.86(1)$ and $P_l=0.32(1)$. To note, when performing this normalization we systematically do not account for the statistical variations of the renormalization factors, which will lead to additional (and unaccounted for) uncertainties on the values of the renormalized population data. 
Also, here we attribute the relatively higher lower-bound value compared to that in \cite{chen2024} to the apparent degradation of the coherence of our Rydberg lasers used for STIRAP (this affects the state preparation efficiency, leading to a higher probability for atoms to be left in the ground state). We also note that, given that $P_l$ partly depends on the decay of the Rydberg states, it should maintain at almost the same level when performing measurement for different evolution times, since we only vary the microwave pulse duration embedded in our fixed $5~\mu{\rm s}$ trap release window (with a fixed time between the Rydberg state excitation and de-excitation). Our detection (based on imaging ground state atoms) always happens at a fixed time point in the sequence.

\section{Breakdown of BO for atom pairs indicated by damped $P_0$ dynamics}

An outstanding feature for the breakdown of Bloch oscillations is wavepacket spreading, evidenced by the damping of short-time dynamics of $P_0$, as shown in the main text. Numerically, we perform fittings with a damped sinewave function to the simulated time evolution of $P_0$. In the $V-\Omega$ panel, the breakdown region is indicated by a non-zero damping coefficient; see Fig.~\pref{FIG:figs6}{b}. Three clear regions are apparent, and both the lower and upper boundaries are proportional to $\Omega$, consistent with the analytical form derived from the energy gap analysis. Our experimental result, shown in Fig.~\pref{FIG:figs6}{d}, also demonstrates the localized oscillation without damping (almost perfectly oscillating back to $\ket{0,0}$) outside the breakdown region. Figures~\pref{FIG:figs6}{a,c} further show a clear dependence of the oscillating frequency $\omega$ on $V$. For small $\Omega\to 0$, $\omega$ approches $\pm (V-\Delta)$ [see Eq.(\ref{eqS2})], with extremely narrow breakdown region around $V=\Delta$. An increase of $\Omega$ smooths the sharp turning point of $\omega$ over interaction-to-detuning ratio at $V/\Delta=1$.

\begin{figure*}[]	
    \includegraphics[width=0.5\textwidth]{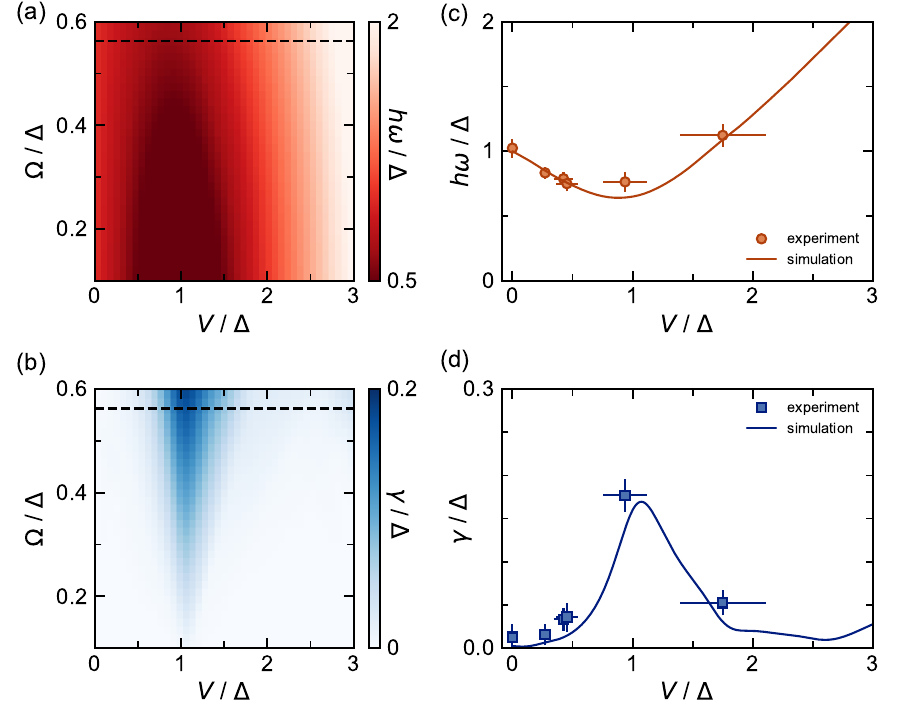}
	\caption{\textbf{Breakdown region of BO for atom pairs from fittings of $P_0$ dynamics.}
        \textbf{(a, b)}~The oscillation frequency $\omega$ (a) and damping coefficient $\gamma$ (b) resolved from fittings of numerically simulated $P_0$ dynamics with a damped sinewave function $P_0(t)=A e^{-\gamma t/\hbar}\cos^2{(\pi \omega t)} + c$, for different interaction strength $V$ and coupling strength $\Omega$. The horizontal dashed lines indicate the $\Omega/\Delta =0.56$ used in the experiment.
        \textbf{(c, d)}~The oscillation frequency $\omega$ (c) and damping coefficient $\gamma$ (d) for different interaction strength $V$ with $\Omega/\Delta=0.56$ (experimental condition). Solid lines are numerical simulation results.
}
\label{FIG:figs6}
\end{figure*}

\section{Determination of the oscillation frequency and breakdown region for atom pairs}

\begin{figure*}[b]	
    \includegraphics[width=\textwidth]{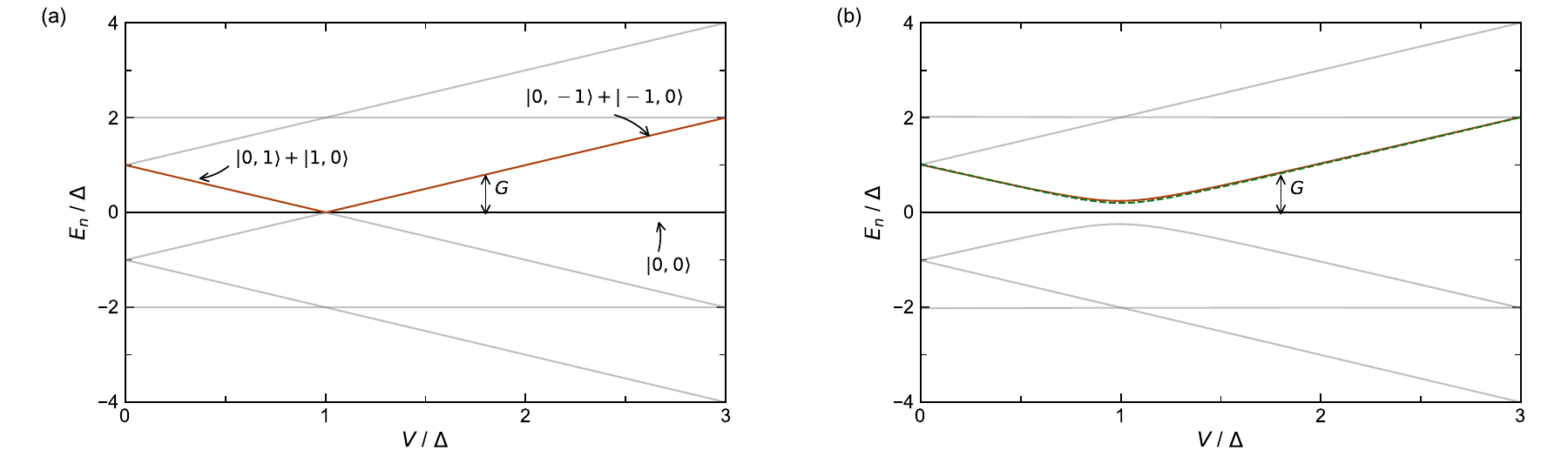}
	\caption{\textbf{Eigenenergy distribution under different interaction strengths for the 3-state subsystem.}
        \textbf{(a)}~With $\Omega=0$, the zero energy band (solid black line) corresponds to the eigenstate $\ket{0,0}$. The eigenstate for the neighboring band (solid orange line) is the pair triplet: $(\ket{0,1}+\ket{1,0})/\sqrt{2}$ for $V<\Delta$ and $(\ket{0,-1}+\ket{-1,0})/\sqrt{2}$ for $V>\Delta$ (and vice-versa for the neighboring negative energy branch). The energy gap between the central bands is labelled as $G = \pm (\Delta -V)$.
        \textbf{(b)}~For finite small $\Omega/\Delta=0.2$, the central bands open up a gap at $V/\Delta$, corresponding to an eigenstate as a superposition of the $\ket{0,0}$ state and pair triplets. The green dashed line indicates the approximated formula $G = \sqrt{|V-\Delta|^2+\Omega^2}$. All gray lines in (a) and (b) correspond additional energy bands for this truncated system.
}
\label{FIG:figS1}
\end{figure*}

Here we show how the interactions affect the oscillation frequency for atom pairs, and additionally we estimate the region in which interactions lead to a breakdown of localization effects. For large detunings, e.g., $\Delta>\Omega$ as used in Fig.~3 in the main text, the non-interacting Bloch oscillation is almost restricted to the center three sites. In this large-bias regime, the oscillation frequency is approximately equal to the energy gap between $\ket{0}$ and $\ket{\pm 1}$ under the dressed state picture. That is, when the dipolar exchange interaction is introduced, the atom pair in $\ket{0,0}$ undergoes collective coupling to the neighboring triplet states (or superpositions of multiple triplets if we consider beyond the truncated $\ket{-1}$, $\ket{0}$, $\ket{1}$ system). Then the oscillation frequency is determined by the energy gap between $\ket{0,0}$ and the neighboring eigenstates. 

Instead of considering the whole 9-state system, we restrict ourselves to the 3-state subsystem, $\{\ket{0}, \ket{\pm 1}\}$, to get the approximate relation between the oscillation frequency and the energy gap. The Hamiltonian is
\begin{equation}
 H = \pm\Delta \ket{\pm 1}\bra{\pm 1}+\frac{\Omega}{2}\left(\ket{0}\bra{\pm 1} + {\rm h.c.}\right) + V_{0,\pm 1}\left(\ket{0, \pm 1}\bra{\pm 1, 0} + {\rm h.c.}\right)~
\end{equation}
with $V_{0,\pm 1} = \mp V$.
We first consider the case with $\Omega \to 0$, for which the eigenenergy distribution under different interaction strengths is shown in Fig.~\pref{FIG:figS1}{a}. The initial $\ket{0,0}$ state is the eigenstate of the zero energy band, while the eigenstates for the neighboring bands with energy $E_n=\pm (V-\Delta)$ are pair triplet states. The atom pair should in principle oscillate with a frequency given by the energy difference of the two bands, i.e., $G=\pm (V-\Delta)$ (although there would be zero amplitude of oscillation in the $\Omega \to 0$ limit). This simple prediction relates to the dashed line shown in Fig.~3(c) of the main text. Small but finite $\Omega$ introduces perturbations to the energy bands and opens up a gap at $V/\Delta=1$; see Fig.~\pref{FIG:figS1}{b}. Now the exact formula of the energy gap between the bands reads $G = \frac{1}{2} \{10 \Delta^2-4\Delta V+2 V^2+5 \Omega^2-[(6 \Delta^2+4 \Delta V-2 V^2 + \Omega^2)^2 + 8\Omega^2 (3 \Delta^2-10 \Delta V+3 V^2+\Omega^2)]^{1/2}\}^{1/2}$. For small $\Omega$, we can neglect the second $8\Omega^2$ term under the inner square root and get the approximated form
\begin{equation}\label{eqS2}
 G \approx \sqrt{|\Delta-V|^2 + \Omega^2},
\end{equation}
as shown by the dotted line in Fig.~3(c) of the main text. We can see from Fig.~\pref{FIG:figS1}{b} that this approximation works well.

Additionally, we can estimate the parameter region over which this approximated form should be invalid. Here we determine the breakdown boundaries of the localization and pair oscillations by letting the energy gap between the two bands be fully covered by the collective pair hopping rate, i.e., 
\begin{equation}
 \sqrt{|\Delta-V|^2 + \Omega^2}=\sqrt{2}\Omega \ . 
\end{equation}
This leads to the simple relation $V=\Delta\pm\Omega$, shown as the two vertical dashed lines in Fig.~3(c) of the maintext. We also validate this formula from our numerical simulations, as the oscillatory fits to the $P_0$ dynamics with collapse in this ``breakdown'' region.

\section{Time evolution of the atom-pair correlations}

\begin{figure*}[b]
	\includegraphics[width=\textwidth]{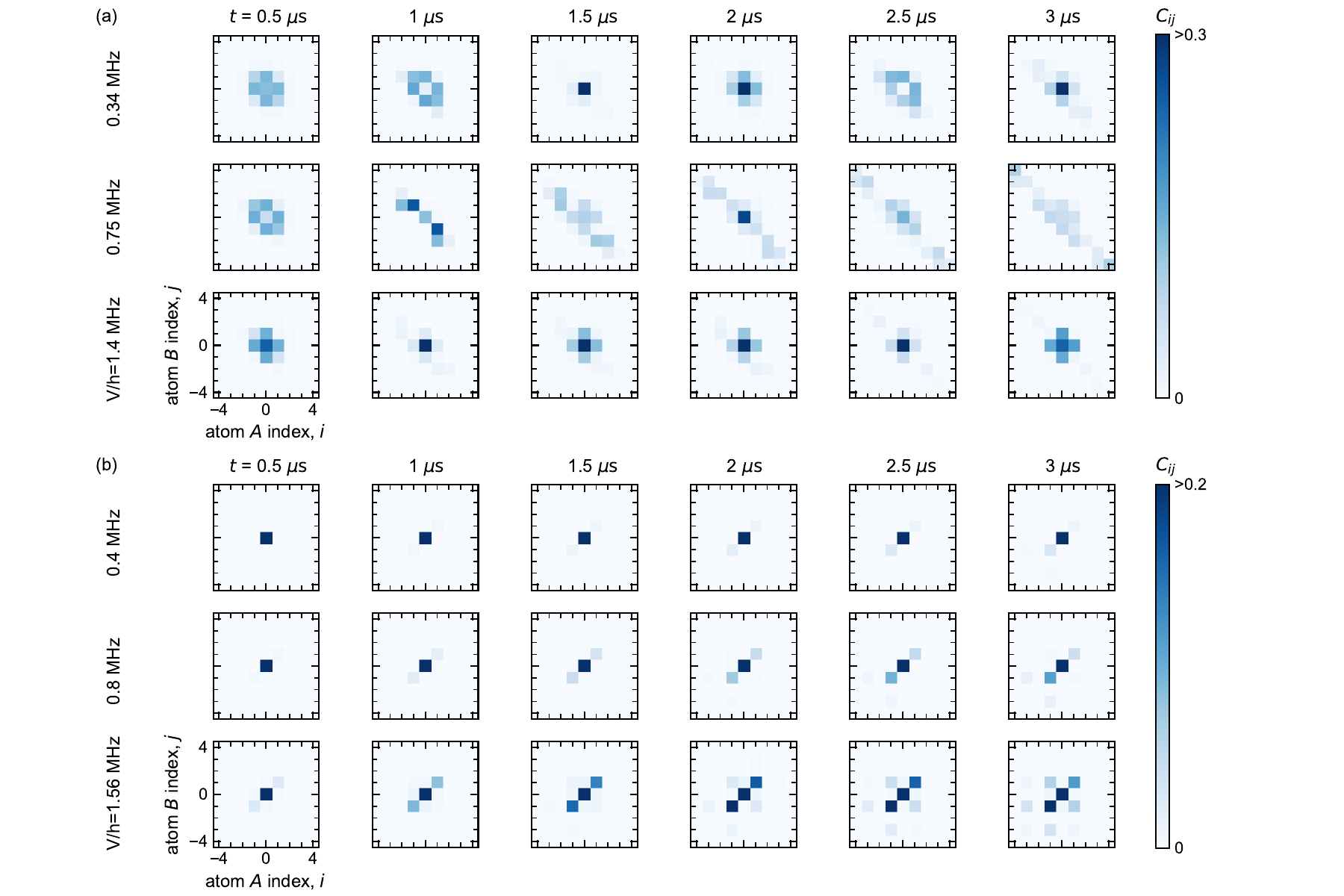}
	\caption{\textbf{Numerically simulated time evolution of the atom pair correlations.}
        \textbf{(a)}~Pair correlation $C_{ij}$ after different evolution times $t$ for the tilted 9-state lattice depicted in Fig.~1(a) of the main text [i.e., same conditions as in Fig.~3 of the main text]. Here $\Delta/h = 0.8~{\rm MHz}$ and $\Omega/h=0.45~{\rm MHz}$.
        \textbf{(b)}~Pair correlation $C_{ij}$ after different evolution times $t$ for the 9-state lattice driven by bichromatic microwaves, as depicted in Fig.~4(a) of the main text. Here $\Delta/h = 5.0~{\rm MHz}$ and $\Omega/h=0.90~{\rm MHz}$.
}
\label{FIG:figS4}
\end{figure*}

Figure~\ref{FIG:figS4} shows the time evolution of the atom-pair correlation $C_{ij}=\langle c^\dagger_{i,A} c^\dagger_{j,B}c_{i,A} c_{j,B} \rangle$ for two of the quantum walk situations discussed in the main text - the case of two interacting particles hopping in a static, tilted lattice (the $C_{ij}$ plots of Fig.~\pref{FIG:figS4}{a}, corresponding to the scenario explored in Fig.~3) and the case of two interacting particles hopping in a tilted lattice under bichromatic driving (the $C_{ij}$ plots of Fig.~\pref{FIG:figS4}{b}, corresponding to the scenario explored in Fig.~4). In these two respective scenarios, the dynamics of the $C_{ij}$ distributions reveal the anti-correlated and correlated nature of the hopping of the two particles (Rydberg electrons) in the synthetic dimension.

In the case of Fig.~\pref{FIG:figS4}{a}, the anti-correlated dynamics of the two interacting particles can be understood simply from a consideration of energy conservation. In this context, it is helpful to consider the $C_{ij}$ graphs as plotting the density dynamics of pair states $\ket{i}\ket{j}$ in an effective two-dimensional lattice. The energies of these pair states represent an effective potential landscape. In the case where atoms $A$ and $B$ both experience a tilted lattice, the effective pair state potential energy landscape has a gradient along the diagonal direction ($i = j$), with lines of equal-energy states along the anti-diagonal ($i = -j$). The single-particle hopping terms act separably as hopping in the $i$ and $j$ directions. In the absence of interactions, because of this separability of the Hamiltonian along $i$ and $j$, the atoms independently experience Stark localization in the presence of a tilted lattice. Dipolar exchange interactions, having the form $\ket{i}\ket{j} \leftrightarrow \ket{j}\ket{i}$, introduce effective hopping in this two-dimensional pair-state lattice that is not separable along $i$ and $j$. Thus, interactions between the atoms facilitate the breakdown of Stark localization, essentially allowing the atom pair states to delocalize along the resonant anti-diagonal channel. This pair state delocalization along the $i = -j$ direction naturally leads to the buildup of anti-correlations. When the the interactions $V$ become sufficiently strong, pair dynamics along the resonant channel become disrupted. This suppression is also easy to understand: the initial $\ket{0}\ket{0}$ state is only (by symmetry) connected to the triplet states $(\ket{-1}\ket{0} + \ket{0}\ket{-1})/\sqrt{2}$ and $(\ket{1}\ket{0} + \ket{0}\ket{1})/\sqrt{2}$. Each of these states experiences a large interaction shift relative to the non-interacting $\ket{0}\ket{0}$ state, such that direct dynamics are arrested for $V \gg \Omega$.

The dynamics seen in Fig.~\pref{FIG:figS4}{b}, under a strong lattice tilt and bichromatic driving, reveal a contrasting buildup of mostly positive (along the diagonal) $C_{ij}$ correlations. This is consistent with the microscopic derivation and description of activated pair-hopping (Eq.~\ref{pairhopp}), and its generalization to the multi-state driven lattice. As can be seen in Fig.~\pref{FIG:figS4}{b}, the buildup of correlations is not purely along the diagonal. This imperfect correlation stems from the presence of additional processes (e.g., $\ket{0}\ket{0} \leftrightarrow \ket{0}\ket{2}$) as discussed below, which are allowed because of the equal $\Delta$ values applied along adjacent links. As discussed below, such processes can in principle be suppressed, in which case one would expect purely potitive (diagonal) correlations to build up in the strongly tilted and driven lattice.

\section{Perturbation theory for the pair hopping rate in the bichromatic driving field}

\begin{figure*}[t]
	\includegraphics[width=\textwidth]{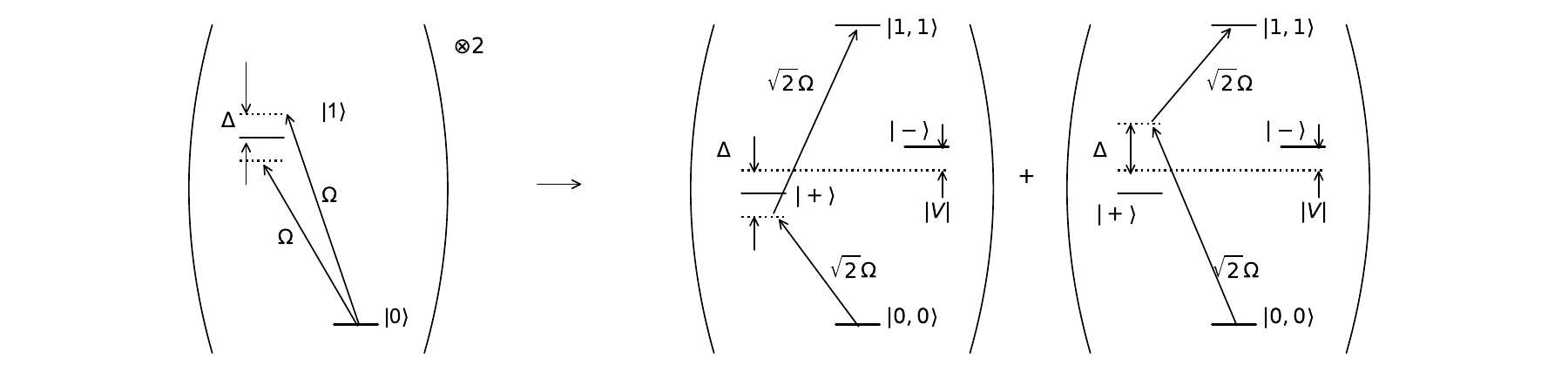}
	\caption{\textbf{Decomposition of the pair hopping under the two-color microwave field into two 2-photon transitions that operate via the intermediate triplet state $|+\rangle$.}
    To note, the $\ket{-}$ singlet state is decoupled from the initial $\ket{0,0} \equiv \ket{0}\ket{0}$ state due to the symmetry of the global driving fields. The processes depicted at right are what give rise to the form of the effective pair-hopping term presented in Eq.~\ref{pairhopp}.
}
\label{FIG:figS2}
\end{figure*}

In large $\Delta$ limit, i.e., $\Delta \gg \Omega, |V|$, the pair oscillation $\ket{0,0} \leftrightarrow \ket{1,1}$ under the bichromatic microwave field is activated by two resonant 2-photon transitions via  the intermediate triplet state, as shown in Fig.~\ref{FIG:figS2}. By directly applying second-order perturbation theory to the two processes, with the collective Rabi frequency $\sqrt{2}\Omega$ and the single-photon detuning $\Delta\pm|V|$ respectively, we have the effective Rabi frequency for the $\ket{0,0} \leftrightarrow \ket{1,1}$ pair transition
\begin{equation}
    \Omega_{\rm eff} = \frac{(\sqrt{2}\Omega)^2}{2(\Delta - |V|)} + \frac{(\sqrt{2}\Omega)^2}{2(\Delta + |V|)} = \frac{2|V|\Omega^2}{\Delta^2-V^2}.
    \label{pairhopp}
\end{equation}
This is the form used in Fig.~4(c). To note, the summed microwave phase of the two driving fields (both defined with the same real parameter $\Omega$ in Fig.~\ref{FIG:figS2}, but able to take controlled complex phases as well) can impart a controlled hopping phase for this correlated hopping process. And, as noted in the main text, this pair-hopping can be controlled entirely independently of the resonant hopping of individual atoms.

When considering how this effective pair hopping rate generalizes when more states (beyond $\ket{0}$ and $\ket{1}$) are included (as in Fig.~4(d,e) of the main text), two effects become apparent. First, when all of the $\Delta$ applied at the different transitions are set to be equal, additional resonant processes are in principle also allowed, such as $\ket{0}\ket{0} \leftrightarrow \ket{0}\ket{2}$. This type of process is responsible for the fact that the pair correlations that develop in Fig.~\pref{FIG:figS4}{b} do not lie purely along the diagonal ($i=j$). Second, one finds that, because the relevant exchange interactions $V$ along different links vary along the synthetic dimension, and given the scaling of $\Omega_{\rm eff}$  with $V$, one expects that in general the pair hopping along a synthetic lattice will be quite non-uniform, and nearly disordered.

Importantly, the ability to control the parameters for each site-to-site link affords enough flexibility to both (1) suppress processes that are not of the form $\ket{i}\ket{i} \leftrightarrow \ket{i+1}\ket{i+1}$ by setting unique (or simply staggered) values of $\Delta$ and (2) compensate for non-uniformities of the $V$ and $\Delta$ values by locally controlling the $\Omega$ terms to achieve uniform values of the $\Omega_{\rm eff}$ pair-hopping rates across each link.

\section{Hamiltonian for the 9-state lattice under bichromatic driving}

Under bichromatic driving, no unitary transformation can be directly applied to obtain an effective time-independent Hamiltonian to describe our system in the full basis of states. We perform numerical simulations with the original time-dependent single particle Hamiltonian
\begin{equation}
    H_{\rm sp} = \sum_j \epsilon_j c_j^\dagger c_j + \frac{\Omega}{2}\sum_{j}\left[\left(e^{-i(\omega_{j}+\Delta)t/\hbar}+e^{-i(\omega_{j}-\Delta)t/\hbar}\right)c_j^\dagger c_{j+1}+{\rm h.c.}\right]
\end{equation}
where $\epsilon_j$ is the potential energy for site $\ket{j}$, $\omega_j=\epsilon_{j+1}-\epsilon_j$ the on-resonant energy for transition $\ket{j}\leftrightarrow\ket{j+1}$, and $\Omega$, $\Delta$ are the Rabi coupling strength and the energy tilting, respectively. We also include the interacting Hamiltonian $H_{\rm int}$, Eq.(2) in the main text, for atom pair dynamics. In practice, since $\omega_j\sim h\times50~{\rm GHz}$ dominates the energy scale, we set the time step as small as possible when integrating the time-dependent Hamiltonian.

\section{Long-time pair dynamics for the 9-state lattice under bichromatic driving with weak interactions}

\begin{figure*}[]
	\includegraphics[width=\textwidth]{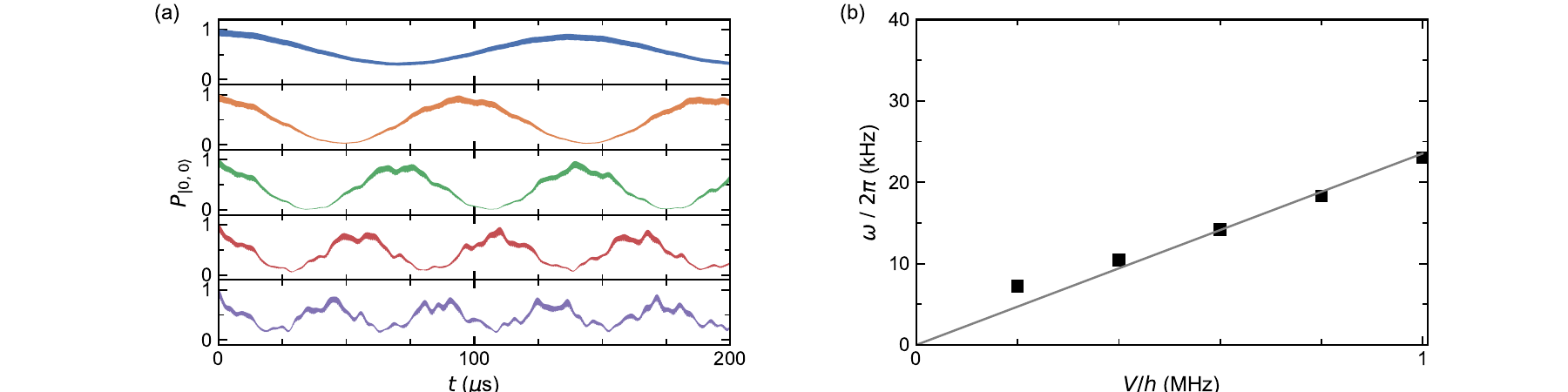}
	\caption{\textbf{Long time $P_{\ket{0,0}}$ dynamics for the 9-state lattice under bichromatic microwave fields.}
        \textbf{(a)}~Numerically calculated time evolution of the population in $\ket{0,0}$ in the 9-state lattice with the bichromatic driving scheme of Fig.~4(a) in the main text. From top to bottom, the interaction strengths are $V/h=\{0.2, 0.4, 0.6, 0.8, 1.0\}~{\rm MHz}$. Here we use $\Omega/h=0.90~{\rm MHz}$ and $\Delta/h=5.0~{\rm MHz}$.
        \textbf{(b)}~The fit-determined oscillation frequency of $P_{\ket{0,0}}$, $\omega$, from fits to the dynamics in (a) with a cosine function $P_{\ket{0,0}}(t) = a + \cos(\omega t)$. The solid line shows the approximately linear relationship between the oscillation frequency and the interaction strength $V$.
}
\label{FIG:figS3}
\end{figure*}

In our experiment on the 9-state lattice with each Rydberg state pair driven by a bichromatic field, we only measure the short time dynamics due to the short trap-release-recapture time window (less than 10~$\mu$s) and finite Rydberg lifetime.
Here we show in Fig.~\pref{FIG:figS3}{a} the long-time dynamics from numerical simulations,
merely to justify the functional form of the fit function used to extract the short-time decay coefficients in Fig.~4(d,e) of the main text.
In the weak interaction regime, the population in $\ket{0,0}$ undergoes slow, but periodic oscillations with an oscillation frequency $\omega$ that is nearly proportional to the interaction strength $V$; see Fig.~\pref{FIG:figS3}{b}. 

On the short experimentally relevant timescales (over just a few $\mu$s), where $\omega t \to 0$ for kHz-scale $\omega$, we see that the relevant fitting function $P_{\ket{0,0}} = a+b\cos(\omega t)$ can be approximated as $(a-b) + 2b \exp(-\omega^2 t^2/4)$. In the main text, we fit the short time dynamics with the formula $P_{\ket{0,0}} = c+de^{-\beta t^2}$. By comparing these two forms, we see that $\beta \sim \omega^2/4$. Since $\omega \propto V$, we arrive at the simple approximated relationship $\beta\propto V^2$, as shown in Fig.~4(e) of the main text.

%\bibliographystyle{apsrev4-1}
%\bibliography{intBO}
%\begin{references}
%    \bibitem[28]{chen2024} T. Chen, C. Huang, I. Velkovsky, K. R. A. Hazzard, J. P. Covey, and B. Gadway, {Nat. Commun.} {\bf 15}, 2675 (2024).
%    \bibitem[29]{Mueller-Escher} E. J. Mueller, {Phys. Rev. A} {\bf 70}, 041603 (2004).
%\end{references}

\clearpage
\end{widetext}

\end{document}